\documentclass[a4paper]{article}
\usepackage{amsfonts}

\usepackage{epsfig}
\usepackage{amssymb}
\usepackage{amsmath}
\usepackage{amsthm}
\usepackage{latexsym}
\usepackage{graphicx}
\def\be{\begin{equation}}
\def\ee{\end{equation}}
\def\bc{\begin{center}}
\def\ec{\end{center}}

\title{Parallel retrieval of correlated patterns}
\author{Elena Agliari  \footnote{
    Universit\`a di Parma, Dipartimento di Fisica and INFN Gruppo di Parma, Italy}, Adriano Barra  \footnote{
    Sapienza Universit\`a di Roma, Dipartimento di Fisica and GNFM Gruppo di Roma,  Italy}, Andrea De Antoni
    \footnote{Sapienza Universit\`a di Roma, Dipartimento di Matematica, Italy}, Andrea Galluzzi  \footnote{
    Sapienza Universit\`a di Roma, Dipartimento di Fisica, Italy}}

\begin{document}
\maketitle

\begin{abstract}
In this work, we first revise some extensions of the standard Hopfield model in the low storage limit, namely the correlated attractor case and the multitasking case recently introduced by the authors. The former case is based on a modification of the Hebbian prescription, which induces a coupling between consecutive patterns and this effect is tuned by a parameter $a$. In the latter case, dilution is introduced in pattern entries, in such a way that a fraction $d$ of them is blank.
Then, we merge these two extensions to obtain a system able to retrieve several patterns in parallel and the quality of retrieval, encoded by the set of Mattis magnetizations $\{ m^{\mu}\}$, is reminiscent of the correlation among patterns. By tuning the parameters $d$ and $a$, qualitatively different outputs emerge, ranging from highly hierarchical, to symmetric.
The investigations are accomplished by means of both numerical simulations and statistical mechanics analysis, properly adapting a novel technique originally developed for spin glasses, i.e. the Hamilton-Jacobi interpolation, with excellent agreement. Finally, we show the thermodynamical equivalence of this associative network with a (restricted) Boltzmann machine and study its stochastic dynamics to obtain even a dynamical picture, perfectly consistent with the static scenario earlier discussed.
\end{abstract}

\section{Introduction} \label{sec:intro}

In the past century, the seminal works by Minsky and Papert \cite{Minsky-1969book}, Turing \cite{Turing-1950Mind}  and von Neumann \cite{neumann-book} set the basis of modern artificial intelligence and, remarkably, established a link between robotics and information theory \cite{Kinchin-1957book}. Another fundamental contribution in this sense was achieved by Hopfield \cite{hopfield-pnas}, who, beyond offering a simple mathematical prescription for the Hebbian rule for learning \cite{hopfield-pnas}, also
pointed out that artificial neural networks can be embedded in a statistical mechanics framework. The latter was rigorously settled by Amit, Gutfreund and Sompolinsky  (AGS) \cite{Amit-2003book}, ultimately reinforcing the bridge between cybernetics and information theory \cite{Jaynes-1957PR}, given the deep connection between the latter and statistical mechanics \cite{Kinchin-1949book,Kinchin-1957book}.

As a second-order result, artificial intelligence, whose development had been mainly due to mathematicians and engineers, became accessible to theoretical physicists too: in particular, when Hopfield published his celebrated paper, the statistical mechanics of disordered systems (mainly spin glasses \cite{Mezard-1987book}) had just reached its maturity and served as a  theoretical laboratory where AGS, as well as many others, gave rise to the mathematical backbone of these associative networks.

In a nutshell, the standard Hopfield model can be described by a two-bodies mean-field Hamiltonian (a Liapounov cost function \cite{Amit-2003book}), which somehow interpolates between the one describing ferromagnetism, already introduced by Curie and Weiss (CW) \cite{Barra-2008JSP}, and the one describing spin-glasses developed by Sherrington and Kirkpatrick (SK) \cite{Mezard-1987book}. Its dichotomic variables (initially termed ``spins" in the original CW or SK theories) are here promoted to perform as binary neurons (some "on/off" exasperations of more standard integrate-and-fire models \cite{Kistler-2002book}) and the interaction matrix (called synaptic matrix in this context) assumes a (symmetrized) Hebbian fashion where information, represented as a set of patterns (namely vectors of $\pm 1$ random entries), is stored. One of the main goals achieved by the statistical mechanics analysis of this model is a clear picture where memory is no longer thought of as statically stored into a confined region (somehow similar to hard disks), but it is spread over the non-linear retroactive synaptic loops merging neurons themselves. Furthermore, it has been offering a methodology where puzzling questions, such as the memory capacity of the network or its stability under the presence of noise, could finally be consistently formulated.

The success of the statistical-mechanics analysis of neural networks is confirmed by the fact that several variations on theme followed and many scientific journals dedicated to this very subject arose.
%
For instance, Amit, Cugliandolo, Griniatsly and Tsodsky \cite{Griniasty-1993NeurComp,Cugliandolo-1993NeurComp,Cugliandolo-1994JPA} considered a simple modification of the Hebbian prescription, able to capture the spatial correlation between attractors observed experimentally as a consequence of a proper learning. More precisely, a scalar correlation parameter $a$ is introduced and when its value overcomes a threshold (whose value contains valuable physics as we will explain), the retrieval of a given pattern induces the simultaneous retrieval of its most-correlated counterparts, in some hierarchical way, hence bypassing the standard single retrieval of the original framework (the so called ``pure state").

In another extension, proposed  by some of the authors of the present paper \cite{Agliari-2012lett,Agliari-2012sub}, the hypothesis of strictly non-zero pattern entries is relaxed in such a way that a fraction $d$ of entries is blank. This is shown to imply retrieval of a given pattern without exhausting all the neurons and, following thermodynamic prescriptions (free energy minimizations), the remaining free neurons arrange cooperatively to retrieve further patterns, again in a hierarchical fashion. As a result, the network is able to perform a parallel retrieval of uncorrelated patterns.

Here we consider an Hopfield network exhibiting both correlated patterns and diluted pattern entries, and we study its equilibrium properties through statistical mechanics and Monte Carlo simulations, focusing on the low-storage regime. The analytical investigation is accomplished through a novel mathematical methodology, i.e, the Hamilton-Jacobi technique (early developed in \cite{Guerra-2001FIC,Barra-2008JSP,Genovese-2009JMP}), which is also carefully explained. The emerging behavior of the system is found to depend qualitatively on $a$ and on $d$, and we can distinguish different kinds of fixed points, corresponding to the so called pure-state or to hierarchical states referred to as ``correlated'' , ``parallel'' or ``dense''. In particular, hierarchy among patterns is stronger for small degree of dilution, while at large $d$ the hierarchy is smoother.


Moreover, we consider the equivalence between the Hopfield model and a class of Boltzmann machines \cite{Bengio-2009ML} developed in \cite{Barra-2010JSP,Barra-2012NN} and we show that this equivalence is rather robust and can be established also for the correlated \emph{and} diluted Hopfield studied here. Interestingly, this approach allows  the investigation of dynamic properties of the model which are as well discussed.


The paper is organized as follows. In section $2$, starting from the low-storage Hopfield model, we revise, quickly and pedagogically, the three extensions (and relative phase diagrams) of interest, namely the high storage case (tuned by a scalar parameter $\alpha$), the correlated case (tuned by a scalar parameter $a$) and the parallel case (tuned by a scalar parameter $d$). In Sec. $3$, we move to the general scenario and we present our main results both theoretically and numerically. Then, in Sec. $4$, we analyze the system from the perspective of Boltzmann machines. Finally, Sec. $5$ is devoted to a summary and a discussion of results. The technical details of our investigations are all collected in the appendices.

\section{ Modelization}\label{sec:model}
Here, we briefly describe the main features of the conventional Hopfield model (for extensive treatment see, e.g., \cite{Amit-2003book,Coolen-2005book}).

Let us consider a network of $N$ neurons. Each neuron $\sigma_i$ can take two states, namely, $\sigma_i = +1$ (fire) and $\sigma_i = -1$ (quiescent). Neuronal states are given by the set of variables $\pmb{\sigma} = (\sigma_1, ... , \sigma_N )$. Each neuron is located on a complete graph and the synaptic connection between two arbitrary neurons, say, $\sigma_i$ and $\sigma_j$, is defined by the following Hebb rule:
\begin{equation} \label{eq:hebb}
J_{ij} = \frac{1}{N} \sum_{\mu=1}^P \xi_i^{\mu} \xi_j^{\mu},
\end{equation}
where $\pmb{\xi}^{\mu}=(\xi_1,...,\xi_N)$ denotes the set of memorized patterns, each specified by a label $\mu=1,...,P$.
The entries are usually dichotomic, i.e., $\xi_i^{\mu} \in \{ +1, -1\}$, chosen randomly and independently with equal probability, namely, for any $i$ and $\mu$,
\begin{equation} \label{eq:pattern_equi}
P(\xi_i^{\mu}) = \frac{1}{2} (\delta_{\xi_i^{\mu}-1} +  \delta_{\xi_i^{\mu}+1}),
\end{equation}
where the Kronecker $\delta_{x}$ equals $1$ iff $x=0$, otherwise it is zero.
Patterns are usually assumed as quenched, that is, the performance of the network is analyzed keeping the synaptic values fixed.

The Hamiltonian describing this system is
\begin{equation} \label{eq:hopfield}
H (\pmb{\sigma , \xi}) = - \sum_{i=1}^N \sum_{i > j=1}^N J_{ij} \sigma_i \sigma_j =  - \frac{1}{2N} \sum_{\substack{i,j=1 \\ j \neq i}}^{N,N}  \sum_{\mu=1}^P \xi_i^{\mu} \xi_j^{\mu} \sigma_i \sigma_j,
\end{equation}
so that the field insisting on spin $i$ is
\begin{equation}
h_i (\pmb{\sigma , \xi})= \sum_{\substack{j=1 \\ j \neq i}}^N J_{ij} \sigma_j.
\end{equation}
The evolution of the system is ruled by a stochastic dynamics, according to which the probability that the activity of a neuron $i$ assumes the value $\sigma_i$ is
\begin{equation} \label{eq:glauber}
P(\sigma_i; \pmb{\sigma , \xi}, \beta ) = \frac{1}{2} [1 + \tanh (\beta h_i \sigma_i)],
\end{equation}
where $\beta$ tunes the level of noise such that for $\beta \to 0$ the system behaves completely randomly, while for $\beta \to \infty$ it becomes noiseless and deterministic; note that the noiseless limit of Eq.~(\ref{eq:glauber}) is $\sigma_i(t+1) = \textrm{sign} \left [ (h_i(t) \right]$.

The main feature of the model described by Eqs.~(\ref{eq:hopfield}) and (\ref{eq:glauber}) is its ability to work as an associative memory. More precisely, the patterns are said to be memorized if each of the network configurations $\sigma_i = \xi_i^{\mu}$ for $i = 1, ..., N$, for everyone of the $P$ patterns labelled by $\mu$, is a fixed point of the dynamics. Introducing the overlap $m^{\mu}$ between the state of neurons $\pmb{\sigma}$ and one of the patterns $\pmb{\xi}^{\mu}$, as
\begin{equation} \label{eq:overlap}
m^{\mu} = \frac{1}{N} (\pmb{\sigma \cdot \xi^{\mu}}) = \frac{1}{N} \sum_{i}^N \sigma_i \xi_i^{\mu},
\end{equation}
a pattern $\mu$ is said to be retrieved if, in the thermodynamic limit, $m^{\mu} = \mathcal{O}(1)$.

Given the definition (\ref{eq:overlap}), the Hamiltonian (\ref{eq:hopfield}) can also be written as
\begin{equation} \label{eq:hopfield2}
H (\pmb{\sigma , \xi}) = - N \sum_{\mu=1}^P (m^{\mu})^2 + P = -  N \pmb{m}^2 + P,
\end{equation}
and, similarly,
\begin{equation}\label{eq:campo2}
h_i (\pmb{\sigma , \xi}) = \sum_{\mu=1}^P \xi_i^{\mu} m^{\mu} - \frac{P}{N}\sigma_i.
\end{equation}

The analytical investigation of the system is usually accomplished in the thermodynamic limit $N \rightarrow \infty$, consistently with the fact that real networks are comprised of a very large number of neurons. Dealing with this limit, it  is convenient to specify the relative number of stored patterns, namely $P/N$ and to define the ratio $\alpha = \lim_{N \rightarrow \infty} P/N$. The case $\alpha=0$, corresponding to a number $P$ of stored patterns scaling sub-linearly with respect to the amount of performing neurons $N$, is often referred to as ``low storage''. Conversely, the case of finite $\alpha$ is often referred to as ``high storage''.

\begin{figure}
 \begin{center}
\includegraphics[width=.48\textwidth]{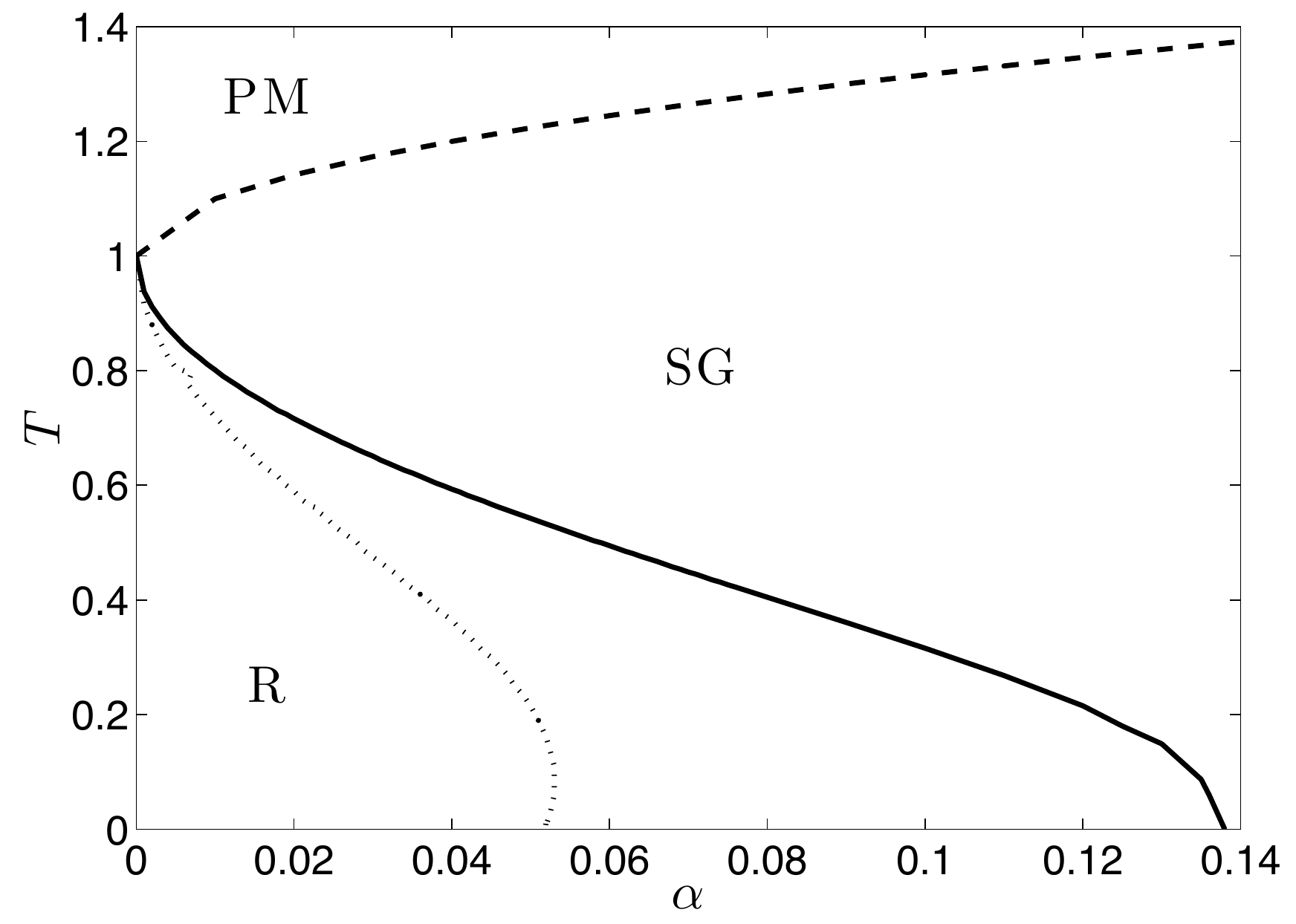}
\caption{\label{fig:Amit} At a high level of noise the system is ergodic (PM) and no retrieval can be accomplished ($m_{\mu}=0, \forall \mu$). By decreasing the noise level below a critical temperature (dashed line) one enters a ``spin-glass'' phase (SG), where there is no retrieval ($m_{\mu}=0$), yet the system is no longer full-ergodic.
Now, if the number of patterns is small enough ($\alpha < 0.138$), by further decreasing the level of noise, one eventually crosses a line (solid curve), below which the system develops $2P$ meta-stable retrieval states, each can be separately retrieved with a macroscopic overlap ($m_{\mu} \neq 0$).
Finally, when $\alpha$ is small enough ($\alpha < 0.05$), a further transition occurs at a critical temperature (dotted line), such that below this line the retrieval states become global minima (R).}
\end{center}
\end{figure}

The overall behavior of the system is ruled by the parameters $T \equiv 1/\beta$ (fast noise) and $\alpha$ (slow noise) and it can be summarized by means of the phase diagram shown in Fig.~\ref{fig:Amit}
Notice that for $\alpha=0$, the so-called pure-state ansatz
\begin{equation}
\pmb{m} = (1,0,...,0),
\end{equation}
always corresponds to a stable solution for $T<1$; the order in the entries is purely conventional and here we assume that the first pattern is the one stimulated.

\section{Generalizations} \label{sec:generalizations}
The Hebbian coupling in Eq.~\ref{eq:hebb} can be generalized in order to include possible more complex combinations among patterns; for instance, we can write
\begin{equation} \label{eq:J_A}
J_{ij} = \frac{1}{N} \sum_{\mu, \nu =1}^{P,P} \xi_i^{\mu} X_{\mu \nu} \xi_j^{\nu},
\end{equation}
where $\pmb{X}$ is a symmetric matrix
; of course, by taking $\pmb{X}$ equal to the identity matrix we recover Eq.~\ref{eq:hebb}.
A particular example of generalized Hebbian kernel was introduced in \cite{Griniasty-1993NeurComp}, and further investigated in \cite{Cugliandolo-1993NeurComp,Cugliandolo-1994JPA}, as
\begin{eqnarray} \label{eq:connection}
\pmb{X}  = \left(
\begin{array}{ccccc}
1 & a & 0 & \cdots & a  \\
a & 1& a & \cdots & 0  \\
\vdots & \vdots & \ddots & \vdots & \vdots  \\
a & 0 & \cdots & a & 1  \\
\end{array}
\right).
\end{eqnarray}
In this way the coupling between two arbitrary neurons turns out to be
\begin{equation}\label{eq:J_leticia}
J_{ij} = \frac{1}{N} \sum_{\mu=1}^P [\xi_i^{\mu} \xi_j^{\mu} + a(\xi_i^{\mu+1} \xi_j^{\mu} +\xi_i^{\mu-1} \xi_j^{\mu} ) ].
\end{equation}
Hence, each memorized pattern, meant as a cyclic sequence, couples the consecutive patterns with a strength $a$, in addition to the usual auto-associative term.

This modification of the Hopfield model was proposed in \cite{Griniasty-1993NeurComp} to capture some basic experimental features about coding in the temporal cortex of the monkey \cite{Chang-1988a,Chang-1988b}: a temporal correlation among visual stimuli can evoke a neuronal activity displaying spatial correlation.
%
%
Indeed, the synaptic matrix (\ref{eq:J_leticia}) is able to reproduce this experimental feature in both low \cite{Griniasty-1993NeurComp,Cugliandolo-1993NeurComp} and high \cite{Cugliandolo-1994JPA} storage regimes.

For the former case,
one derives the mean-field equations determining the attractors, which, since the matrix is symmetric, are simple fixed points. In the limit of a large network, they read off as \cite{Griniasty-1993NeurComp}
\begin{equation} \label{eq:selfcons_a}
m^{\mu} = \left \langle \xi^{\mu} \, \tanh \left( \sum_{\mu=1}^P m^{\mu} [\xi_i^{\mu} + a (\xi_i^{\mu+1} + \xi_i^{\mu-1}) ] \right) \right \rangle_{\xi},
\end{equation}
where $\langle \cdot \rangle_{\xi}$ means an average over the quenched distribution of patterns.

In \cite{Griniasty-1993NeurComp}, the previous equation was solved by starting from a pure pattern state and iterating until convergence.
%
%
In the noiseless case, where the hyperbolic tangent can be replaced by the sign function, the pure state ansatz is still a fixed point of the dynamics if $a \in [0, 1/2)$, while if $a \in (1/2,1]$, the system evolves to an attractor characterized by the Mattis magnetizations (assuming $P \geq 10$, see Appendix A)
\begin{equation} \label{eq:ansatz_leti}
\pmb{m} = \frac{1}{2^7} (77,51,13,3,1,0,...,0,...,0,1,3,13,51),
\end{equation}
namely, the overlap with the pattern used as stimulus is the largest and the overlap with the neighboring patterns in the stored sequence decays symmetrically until vanishing at a distance of $5$.
Some insights into these results can be found in Appendix A.

\begin{figure}
 \begin{center}
\includegraphics[width=.48\textwidth]{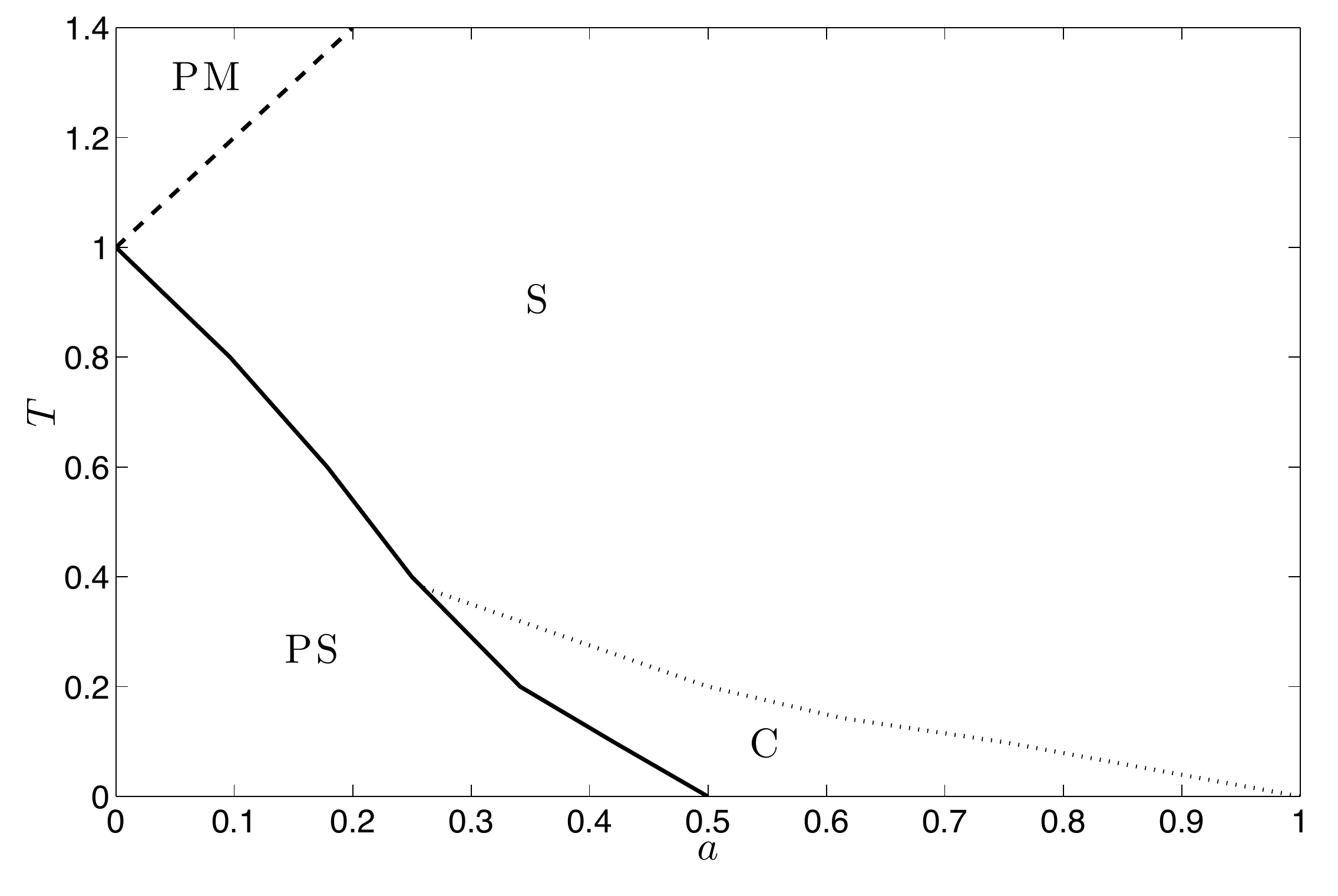}
\caption{\label{fig:Leti} Phase diagram for the correlated model with low storage ($P=13$), as originally reported in \cite{Cugliandolo-1993NeurComp}. At a high level of noise the system is ergodic (PM) and it eventually reaches a state with $m^{\mu}=0, \forall \mu$. At smaller temperatures (below the dashed line), the system evolves to a so-called symmetric state (S), characterized by, approximately, $m^{\mu}=m \neq 0, \forall \mu$.
Then, if $a$ is small enough, by further reducing the temperature (below the solid line), the network behaves as a Hopfield network and the pure state retrieval (PS) can be recovered.
On the other hand, if $a$ is larger, as the temperature is reduced, correlated attractors (C) appear according to Eq.~\ref{eq:ansatz_leti}. Then, if the temperature is further lowered, the system recovers the Hopfield-like regime.
If $a>1/2$, the pure state regime is no longer achievable.
}
\end{center}
\end{figure}

In the presence of noise, one can distinguish four different regimes according to the value of the parameters $a$ and $T$.
%
%
The overall behavior of the system is summarized in the plot of Fig.~\ref{fig:Leti}.
A similar phase diagram, as a function of $\alpha$ and $a$, was drawn in \cite{Cugliandolo-1994JPA} for the high-storage regime.

\bigskip

A further generalization can be implemented in order to account for the fact that the pattern distribution may not be uniform or that pattern may possibly be blank.
For instance, in the latter case one may replace
 Eq.~\ref{eq:pattern_equi} by
\begin{equation} \label{eq:pattern_blank}
P(\xi_i^{\mu}) = \frac{1-d}{2} \delta_{\xi_i^{\mu}-1} +  \frac{1-d}{2} \delta_{\xi_i^{\mu} +1} + d \delta_{\xi_i^{\mu}},
\end{equation}
where $d$ encodes the degree of dilution of pattern entries.
This kind of extension has strong biological motivations, too.
In fact, the distribution in Eq.~\ref{eq:pattern_equi} necessarily implies that
the retrieval of a unique pattern does employ all the available neurons, so that no resources are left for further tasks.
Conversely, with Eq.~\ref{eq:pattern_blank} the retrieval of one pattern still allows available neurons which can be used to recall other patterns. The resulting network is therefore able to process several patterns simultaneously.
The behavior of this system is deeply investigated in \cite{Agliari-2012lett,Agliari-2012sub}, as far as the low storage regime is concerned.

\begin{figure}
 \begin{center}
\includegraphics[width=.48\textwidth]{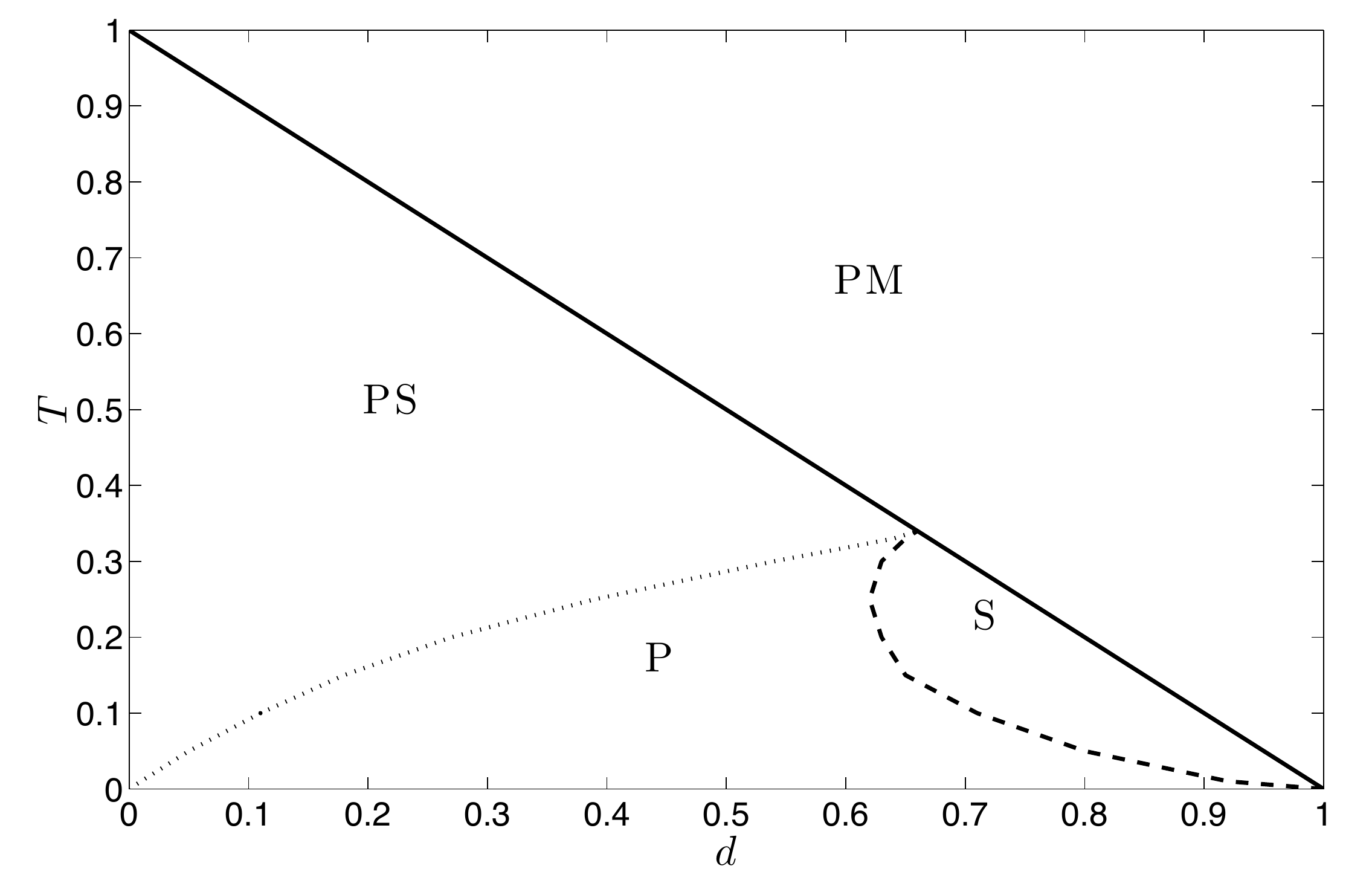}
\caption{\label{fig:diagramma_alps}
At high levels of noise the system is ergodic (PM) and below the temperature $T=1-d$ (continous line) it can develop a pure state retrieval (PS) or a symmetric retrieval (S), according to whether the dilution is small or large, respectively.
At small temperatures and intermediate degree of dilution the system can develop a parallel (P) retrieval, according to Eq.~\ref{eq:h_retrieval}.
The continuous line works for any value of $P$, while the dotted and dashed lines were obtained numerically for the case $P=3$.}
\end{center}
\end{figure}

In particular, it was shown both analytically (via density of states analysis) and numerically (via Monte Carlo simulations), that the system
evolves to an equilibrium state where several patterns are contemporary retrieved; in the noiseless limit $T=0$, the equilibrium state is characterized by a hierarchical overlap
\begin{equation}\label{eq:h_retrieval}
\pmb{m}=(1-d)(1,d,d^2,...,0),
\end{equation}
hereafter referred to as ``parallel ansatz'', while, in the presence of noise, one can distinguish different phases as shown by the diagram in Fig.~\ref{fig:diagramma_alps}.
%


To summarize, both generalizations discussed above, i.e. Eqs.~\ref{eq:J_leticia} and ~\ref{eq:pattern_blank}, induce the break-down of the pure-state ansatz and allow the retrieval of multiple patterns without falling in spurious states \footnote{Since here we focus on the case $\alpha=0$, spurious states are anyhow expected not to emerge since they just appear when pushing the system toward the spin-glass boundary on $\alpha >0$. 
}.
In the following, we merge such generalizations and consider a system exhibiting both correlation among patterns and dilution in pattern entries.


\section{General Case} \label{sec:General}
\begin{figure}
 \begin{center}
\includegraphics[width=.48\textwidth]{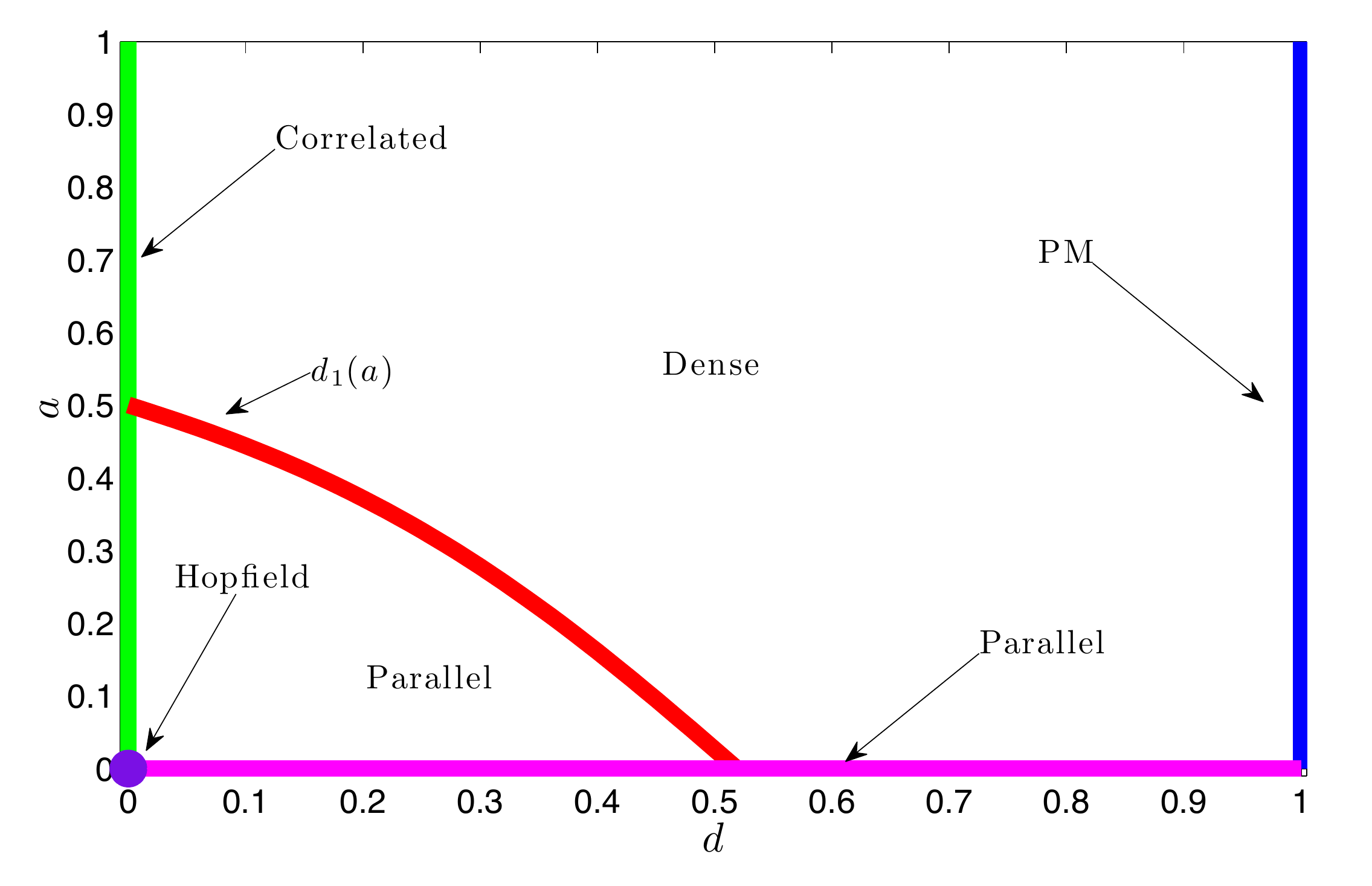}
\caption{\label{fig:square} Schematic representation of the general model considered in the low-stare regime ($\alpha=0$) and zero noise ($T=0$). According to the value of the parameters $a$ (degree of correlation) and $d$ (degree of dilution) the system can recover different kinds of systems. The red curve corresponds to Eq.~\ref{eq:d_1}.}
\end{center}
\end{figure}

Considering a low-storage regime with constant $P$, the general case with $a \in [0,1]$ and $d \in [0,1]$ can be visualized as a square (see Fig.~\ref{fig:square}), where vertices and nodes correspond to either already-known or trivial cases, while the bulk will be discussed in the following.

First, we notice that the coupling distribution is still normal with average $\langle J \rangle_{\xi} =0$ and variance  $\langle J^2 \rangle_{\xi} = (1+2a^2)(1-d)^2/(2P)$. The last result can be realized easily by considering a random walk of length $P$: The walker is endowed with a waiting probability $d$ and at each unit time it performs three steps, one of length $1$ and two of length $a$.


Moreover, as shown in Appendix C, the self-consistance equations found in \cite{Griniasty-1993NeurComp,Agliari-2012lett} can be properly extended to the case $d\neq0$ as
\begin{equation} \label{emmes eq}
\pmb{m} = \langle \pmb{\xi} \tanh \left( \beta \ \pmb{\xi} \cdot  \pmb{X} \pmb{m}\right) \rangle_{\pmb{\xi}},
\end{equation}
where $\pmb{X}$ is the matrix inducing the correlation (see Eq.~\ref{eq:connection}) and the brakets $\langle . \rangle_{\pmb{\xi}}$ now mean an average over the possible realizations of dilution too.
\newline

\subsection{Free-noise system: $T=0$}
\begin{figure}
 \begin{center}
\includegraphics[width=.85\textwidth]{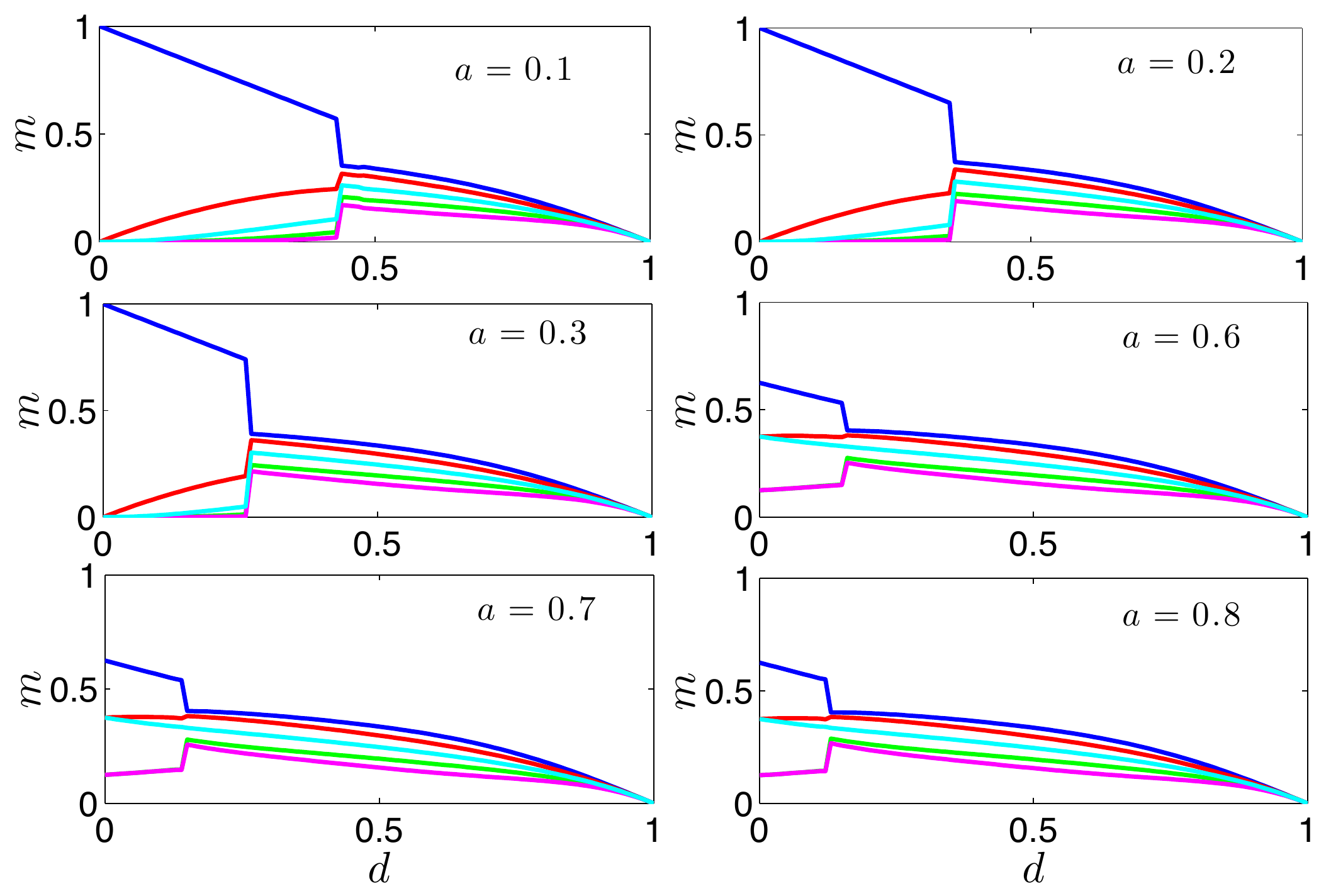}
\caption{\label{fig:P5T0} Magnetization $\pmb{m}$ versus degree of dilution for fixed $P=5$ and $T=0.0001$; magnetizations related to different patterns are shown in different colors. Several values of $a$ are considered, as specified in each panel.}
\end{center}
\end{figure}

\begin{figure}
 \begin{center}
\includegraphics[width=.85\textwidth]{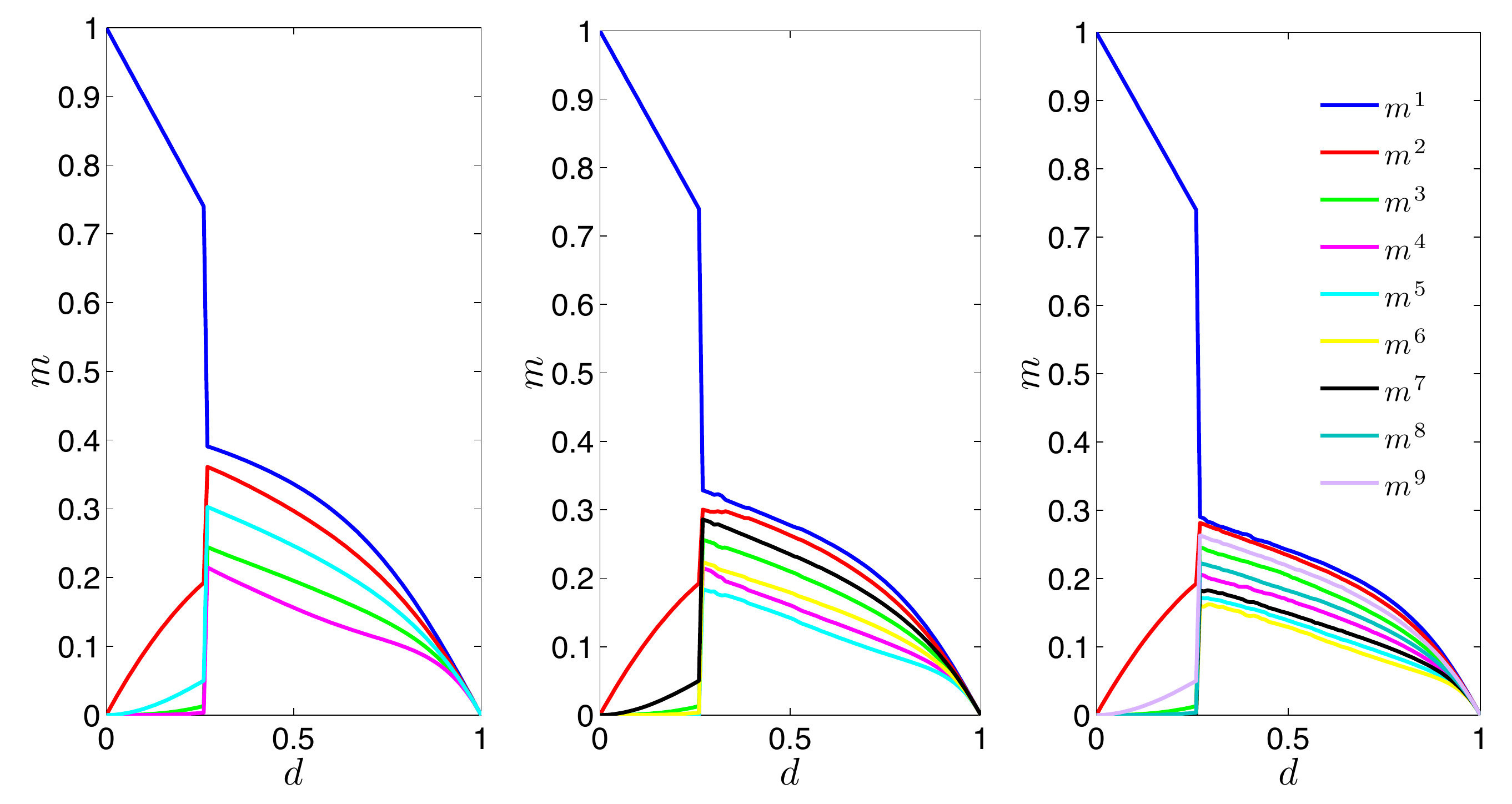}
\caption{\label{fig:P5P7P9T0} Magnetization $\pmb{m}$ versus degree of dilution for fixed $a=0.3$ and $T=0.0001$. Several values of $P$ are considered for comparison: $P=5$ (leftmost panel), $P=7$ (central panel) and $P=9$ (rightmost panel). Magnetizations related to different patterns are shown in different colors.}
\end{center}
\end{figure}

The numerical solution of the self-consistence equation (\ref{emmes eq}) are shown in Figs.~\ref{fig:P5T0} and \ref{fig:P5P7P9T0}, as functions of $d$ and $a$; several choices of $P$ are also compared.
Let us focus on the case $P=5$ (see Fig.~\ref{fig:P5T0}) for a detailed description of the system performance.

When $a<1/2$, the parallel ansatz (\ref{eq:h_retrieval}) works up to a critical dilution $d_1(a)$, above which the gap between magnetizations, i.e. $|m^{\mu} - m^{\nu}|$, drops abruptly and, for $d>d_1(a)$,  all magnetizations are close and decrease monotonically to zero. To see this, let us reshuffle the ansatz in (\ref{eq:h_retrieval}), so to account for the hierarchy induced by correlation, that is,
\begin{equation} \label{eq:ansatz_order}
\pmb{m}=(1-d)(1, d, d^3,d^4,d^2),
\end{equation}
which can be straightforwardly extended to any arbitrary $P$.
Given the state (\ref{eq:ansatz_order}), the field insisting on $\sigma_i$ is
\begin{eqnarray}
h_i &=& \sum_{\mu=1}^P [\xi_i^{\mu} + a (\xi_i^{\mu-1} + \xi_i^{\mu+1}) ] m^{\mu}  \\
\nonumber
&=&(1-d) \{ \xi_i^{1} [1+ad(1+d)] + \xi_i^2 [d+ a (1+d^3)] + \xi_i^3 [d^3 +ad (1+d^3)]\\ \nonumber
&+& \xi_i^4 [d^4 + ad^2 (1+d)] + \xi_i^5 [d^2 + a (1+d^4)] \}.
\end{eqnarray}
A signal-to-noise analysis suggests that this state is stable only for small degrees of dilution. In fact, there exist configurations (e.g., $\xi_i^1 \neq 0$ and $\xi_i^1 = - \xi_i^{\mu}$, for any $\mu>1$) possibly giving rise to a misalignment between $\sigma_i$ and $\xi_i^1$, with consequent reduction of $m^1$.
This can occur only for $d>d_1(a)$, being $d_1(a)$ the root of the equation $a=(1-d-d^2-d^3-d^4)/[2(1+d^3+d^4)]$, as confirmed numerically (see Fig.~\ref{fig:P5T0}). In general, for arbitrary $P$, one has
\begin{equation}
\label{eq:d_1}
a=(1-2d+d^P)/[2(1-d+d^3-d^P)],
\end{equation}
which is plotted in Fig.~\ref{fig:square}.

As $d \geq d_1(a)$, the magnetic configuration corresponding to  Eq.~(\ref{eq:ansatz_order}) undergoes an updating where a fraction of the spins aligned with $\xi^1$ flips to agree mostly with $\xi^2$ and $\xi^5$, and partly also with $\xi^3$ and $\xi^4$; as a result, $m^1$ is reduced, while the other magnetizations are increased. Analogously, a fraction of the spins aligned with $\xi^2$ is unstable and flips so to align mostly with $\xi^5$; consequently, there is a second-order correction which is upwards for $m^5$ (and to less extent for $m^1$, $m^3$ and $m^4$) and downwards for $m^2$. Similar arguments apply for higher-order corrections.

At large values of dilution it convenient to start from a different ansatz, namely from the symmetric state
\begin{equation}
\pmb{m}=\tilde{m}(1, 1, 1, 1,1).
\end{equation}
This is expected to work properly when dilution is so large that the signal on any arbitrary spin $\sigma_i$ stems from only one pattern, i.e., $\xi_i^{\mu} \neq 0$ and $\xi_i^{\nu}=0, \forall \nu \neq \mu$. This approximately corresponds to $d>1 - 1/P$. The related magnetization is therefore $\tilde{m}=d^4(1-d)$. Now, reducing $d$, we can imagine, for simplicity, that each spin $\sigma_i$ feels a signal from just two different patterns, say $\xi^1$ and $\xi^2$. The prevailing pattern, say $\xi^1$, will increase the related magnetization and vice versa. This induces the breakdown of the symmetric condition so that $m^1$ grows larger, followed by $m^2$, $m^5$, and so on. The gap between magnetizations corresponds to the amount of spins which have broken the local symmetry, that is $d^3(1-d)^2$. Thus, magnetizations differs by the same amount and this configuration is stable for large enough dilutions. By further thickening non-null entries, each spin has to manage a more complex signal and higher order corrections arise. For instance, one finds $m^1= d^4(1-d) + 4d^3(1-d)^2 + 2d^2(1-d)^3$, and similarly for $m^{\mu>1}$. This picture is consistent with numerical data and, for large enough values of $d$, it is independent of $a$ (see Fig.~\ref{fig:P5T0}). Notice that in this regime of high dilution hierarchy effects are smoothed down, that is, magnetizations are close and we refer to this kind of state as ``dense''.


When $a>1/2$, the parallel ansatz in Eq.~\ref{eq:ansatz_order} is no longer successful at small $d$, in fact, correlation effects prevail and one should rather consider a perturbed version of the correlated ansatz (\ref{eq:ansatz_leti}), that is,
\begin{equation} \label{eq:ansatz_leti_pert}
\pmb{m}=(1-d)\frac{1}{8}(5, 3, 1,1,3).
\end{equation}
We use (\ref{eq:ansatz_leti_pert}) as initial state for our numerical calculations finding, as fixed point, $m^1=(1-d)5/8$, $m^2=(1-d^2)3/8$, $m^3=m^4 = (1+d)1/8$, $m^5=(1-d+d^2)3/8$.
This state works up to a critical dilution $d_2(a)$, where, again there is the establishment of a situation with magnetizations close and monotonically decreasing to zero. This scenario is analogous to the one describe above and, basically, $d_2(a)$ marks the onset of the region where dilution effects prevails.
The threshold value $d_2$ is slowly decreasing with $a$.




\subsection{Noisy system: $T>0$} The noisy case gives rise to a very rich phenomenology, as evidenced by the plots shown in Fig~\ref{fig:P5T01}.

In the range of temperatures considered, i.e. $T \leq 0.1$, we found that, when $d<d_1(a,T)$ and $a<a_1(T)$, the parallel ansatz (\ref{eq:ansatz_order}) works; in general, $d_1(a,T)$ decreases with $T$ and with $a$, consistently with what found in the noiseless case (see Fig.~\ref{fig:square}).
Moreover, $a_1(T)$ also decreases with $T$, consistently with the case $d=0$ \cite{Cugliandolo-1994JPA}, (see Fig.~\ref{fig:Leti}): from $a_1$ onwards correlation effects get non-negligible. For larger values of $a$, namely $a_1(T) < a < a_2(T)$, the perturbed correlated ansatz (\ref{eq:ansatz_leti_pert}) works, while for $a>a_2(T)$ correlations effects are so important that a symmetric state emerges. Again, we underline the consistentcy with the case $d=0$ \cite{Cugliandolo-1994JPA}: the region $a_1(T) < a<a_2(T)$ corresponds to an intermediate degree of correlation which yields a hierarchical state, while $a>a_2(T)$ corresponds to a high degree of correlation which induces a symmetric state (see Fig.~\ref{fig:Leti}).

As for the region of high dilution, we notice that when $d$ is close to $1$ the paramagnetic state $\pmb{m}=(0,0,0,..,0)$ emerges. In fact, as long as the signal $(1-d) + 2a(1-d)$ is smaller than noise $T$, no retrieval can be accomplished, therefore, the condition
\begin{equation}
d < 1 - T/(1+2a)
\end{equation}
must be fulfilled for $m^{\mu}>0$ to hold. The system then relaxes to a symmetric state which lasts up to intermediate dilution, where a state with ``dense '' magnetizations, analogous to the one described in Sec.~$4.1$, emerges.


\begin{figure}
 \begin{center}
\includegraphics[width=.95\textwidth]{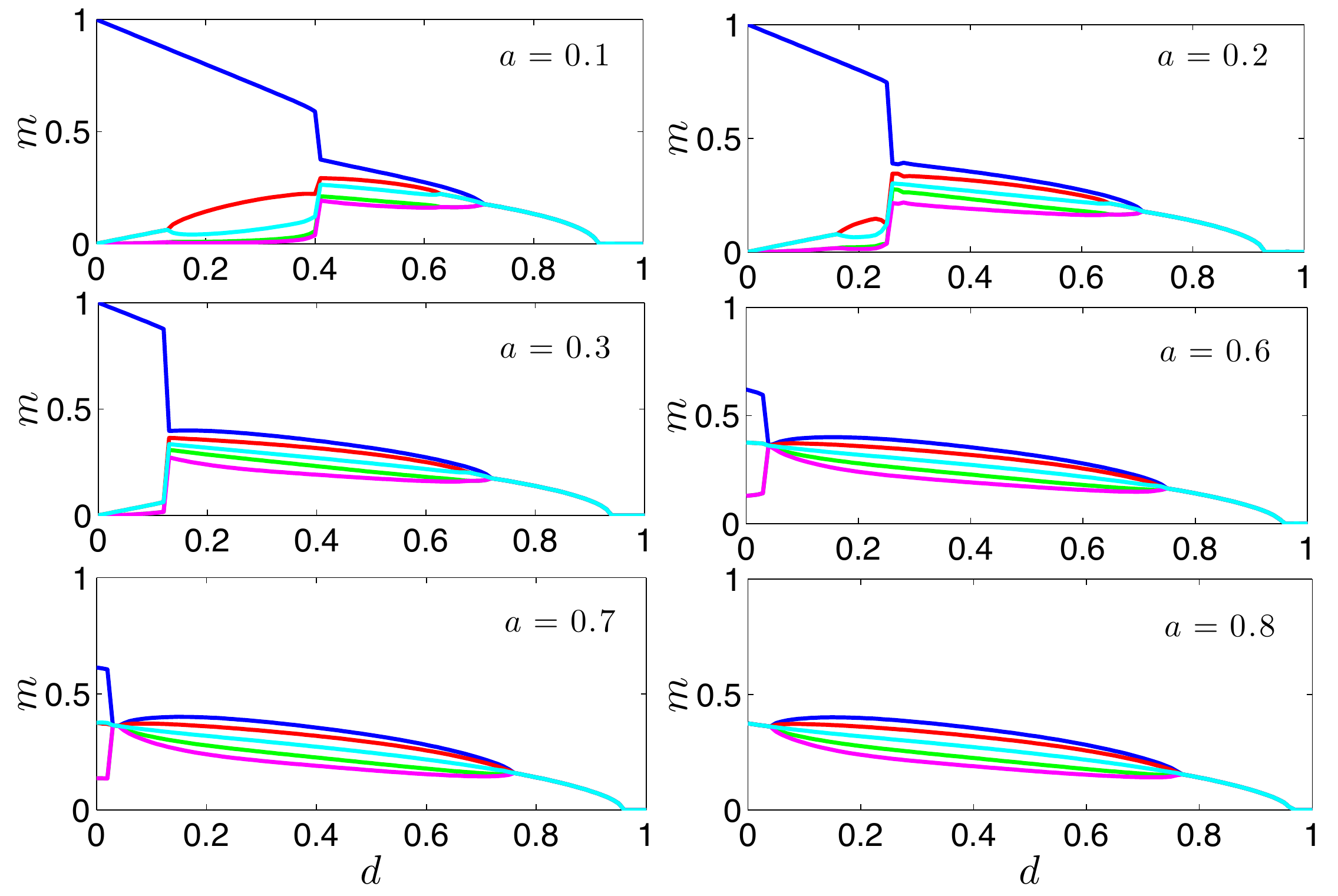}
\caption{\label{fig:P5T01} Magnetization $\pmb{m}$ versus degree of dilution for fixed $P=5$ and $T=0.1$; magnetizations related to different patterns are shown in different colors. Several values of $a$ are considered, as specified in each panel.}
\end{center}
\end{figure}

\subsection{Monte Carlo simulations}
The model was analyzed also via Monte Carlo simulations, which were implemented to determine the equilibrium values of the order parameter associated to the following Hopfield-like Hamiltonian
\be
H=-\sum_{i<j} \sigma_i \sigma_j J_{ij}=\frac{-1}{2N}\sum_{ij} \sigma_i \sigma_j \sum_\mu [\xi_i^\mu \xi_j^\mu+a(\xi_i^{\mu+1} \xi_j^\mu+\xi_i^{\mu-1} \xi_j^\mu)].
\ee
 where the coupling encodes correlation among patterns according to Eq.~(\ref{eq:J_leticia}), and pattern entries are extracted according to Eq.~(\ref{eq:pattern_blank}).

The dynamical microscopic variables evolve under the stochastic Glauber dynamic \cite{Glauber-1963JMP}
\be
\sigma_i(t+\delta t)= \mathrm{sign} [ \tanh[\beta h_i(\pmb{\sigma} (t))]+\eta_i(t)],
\ee
where the fields $h_i=\sum_j J_{ij}\sigma_j(t)$ represent the post-synaptic potentials of the neurons.
The independent random numbers $\eta_i(t)$, distributed uniformly in $[0,1]$, provides the dynamics with a source of stochasticity.
The parameter $\beta=1/T$ controls the influence of the noise on the microscopic variables $\sigma_i$.
In the limit $T \rightarrow 0$, namely $\beta \rightarrow \infty$ the process becomes deterministic and the system evolves according to $\sigma_i(t+\delta t)=\mathrm{sign}[h_i]$.

In general, simulations were carried out using lattices consisting of $10^4$ ``neurons" and averaging on statistical samples composed of $10^2$ realizations.
For each realization of the pattern set $\{ \pmb{\xi}^{\mu} \}_{\mu=1,...,P}$, the equilibrium values of Mattis magnetizations were determined as a function of $d$ and the degree of dilution in pattern entries is incremented in steps of $\Delta d=0.01$, by sequentially set equal to zeros the entries of the $P$ vectors, in agreement with the distribution (\ref{eq:pattern_blank}).

Overall, there is a very good agreement between results from MC simulations, from numerical solution of self-consistent equations and from analytical investigations (see Fig.~\ref{fig:P5T01}).

\begin{figure}
 \begin{center}
\includegraphics[width=.8\textwidth]{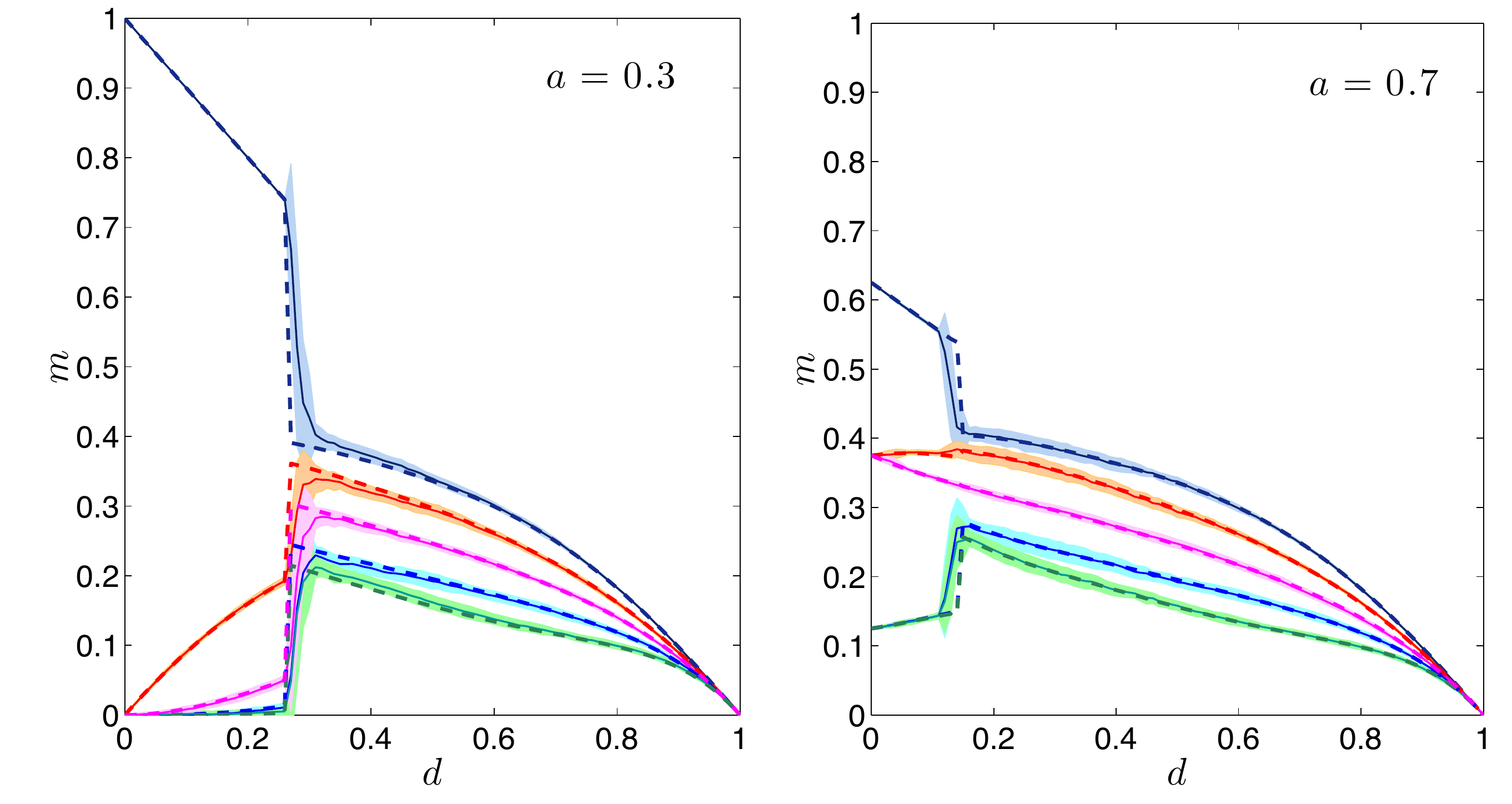}
\caption{\label{fig:P5T01} Magnetization $\pmb{m}$ versus degree of dilution for fixed $P=5$, $T=0.0001$ and $a=0.3$ (left panel) or $a=0.7$ (right panel). Results from numerical solution of Eq.~\ref{emmes eq} (dashed, thick line) and Monte Carlo simulations (solid, thin lines) with associated error (shadows) are compared showing, overall, a very good agreement.}
\end{center}
\end{figure}

\section{Extended Boltzmann machine} \label{sec:BM}
 It is possible to get a deeper insight into the behavior of the system from the perspective of Boltzmann Machines (BMs), exploiting the approach first introduced in \cite{Barra-2012NN}. In particular, it was shown that a  "hybrid" BM
characterized by a bipartite topology (where the two parties are made up by $N$ visible units $\sigma_i$ and by $P$ hidden units $z_{\mu}$, respectively), after a marginalization over the (analog) hidden units, turns out to be (thermodynamically) equivalent to a Hopfield network.   
In this equivalence the $N$ visible units play the role of neurons and the link connecting $\sigma_i$ to $z_{\mu}$ is associated to a weight $\xi_i^{\mu}$.
The term ``hybrid'' refers to the choice of the variables associated to units: the visible units are binary ($\sigma_{i} \in \{ -1, +1 \}$), as in a Restricted Boltzmann Machine, while the hidden ones are analog ($z_{\mu} \in \mathbb{R}$), as in a Restricted Diffusion Network.

\begin{figure}
 \begin{center}
\includegraphics[width=.38\textwidth]{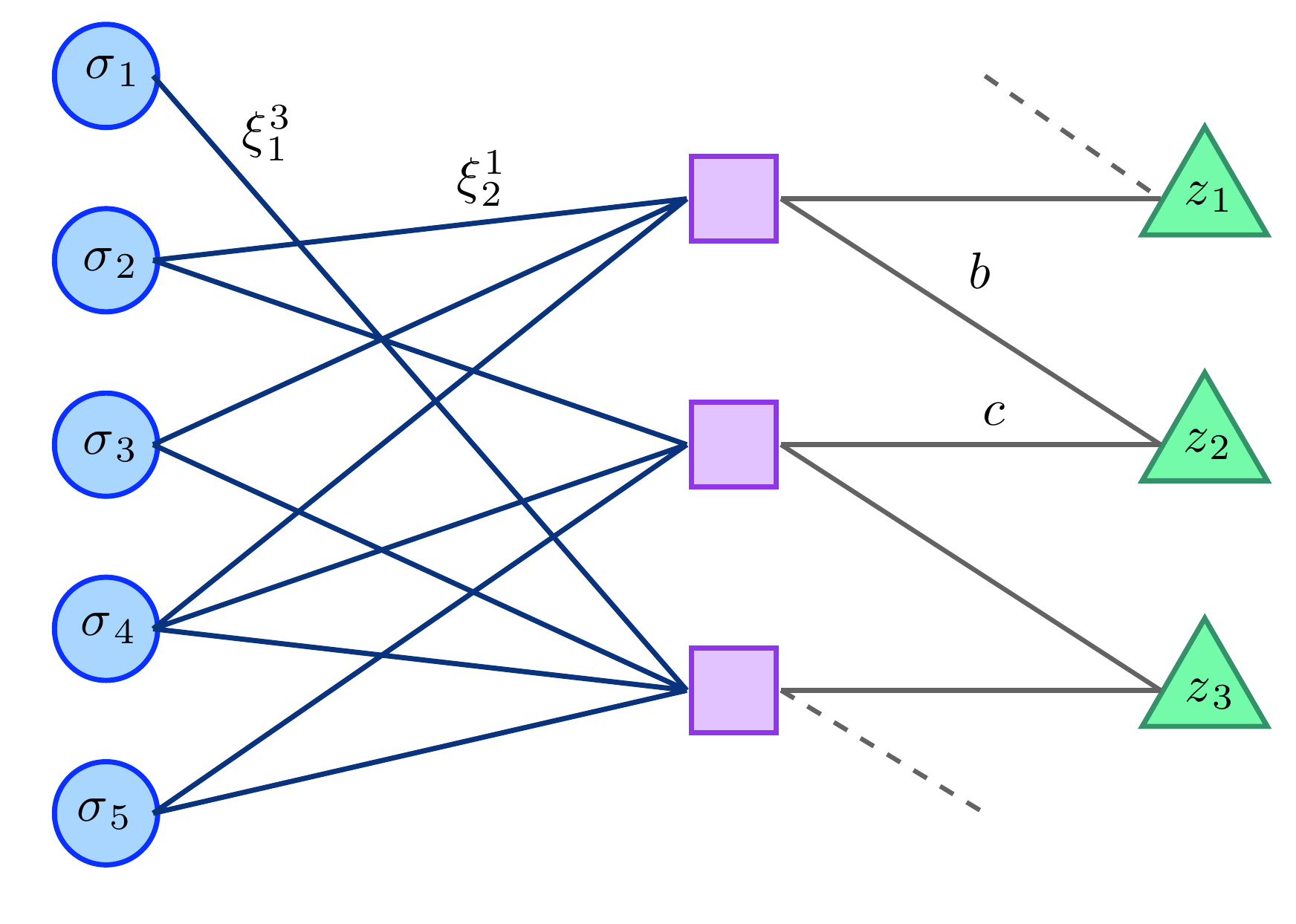}
\caption{\label{fig:BM} Schematic representation of a hybrid BM, with $N=5$ visible nodes ($\bigcirc$) and $P=3$ hidden nodes ($\triangle$). The number of boxes ($\square$) is $P$ as well. The average number of links stemming from visible units is $2$, due to dilution. The link between the $i$-th visible unit and the $\mu$-th box is $\xi_i^{\mu}$; the link between the $\mu$-th box and the $\mu$-th [($\mu+1$)-th] hidden unit is $c$ [$b$].}
\end{center}
\end{figure}

As we are going to show, this picture can be extended to include also the correlation among attractors and the dilution in pattern entries.
More precisely, we introduce an additional layer made up by $P$ ``boxes'', which switches the signal $\xi_i^{\mu}$ on the two hidden variables $z_{\mu}$ and $z_{\mu +1}$ (see Fig.~\ref{fig:BM}).

Such boxes do not correspond to any dynamical variable, but they retain a structural function as they properly organize the interactions between the two ``active'' layers: The binary layer is linked to boxes by a synaptic matrix $\pmb{\xi}$, the boxes are in turn connected to the analog layer by a ``connection matrix'' that we call $\tilde{\pmb{X}}$. The synaptic matrix $\pmb{\xi}$ is $P \times N$ dimensional, each row $\pmb{\xi}^{\mu}$ being a stored pattern. A link between the discrete neuron $\sigma_{i}$ and the $\mu$-th box is drawn with weight $\xi_i^{\mu}$, which take value in the alphabet $\lbrace -1,\ 0,\ 1\rbrace$ following a proper probability distribution. A null weight corresponds to a lack of link, that is, we are introducing a random dilution in the left of the structure. On the other hand, the matrix $\tilde{\pmb{X}}$ is $P \times P$ dimensional and meant to recover the correlation among the stored patterns.
Here, we choose $\pmb{\xi}$ according to Eq.~\ref{eq:pattern_blank} and $\tilde{\pmb{X}}$ such to
recover \cite{Griniasty-1993NeurComp,Cugliandolo-1993NeurComp,Cugliandolo-1994JPA}, namely

\begin{equation} \label{xtilde}
\tilde{X}_{\mu ,\nu}= c \delta_{\mu ,\nu} + b \delta_{\mu ,\nu -1},
\end{equation}
where $c$ and $b$ are parameters tuning the strength of correlation between consecutive patterns entries (vide infra). More complex and intriguing choices of $\tilde{\pmb{X}}$ could be implemented, possibly related to a major adherence to biology.

The dynamics of the hidden and visible layers are quite different.  As explained in \cite{Barra-2012NN}, the activity in the analog layer follows a Ornstein-Uhlembeck (OU) diffusion process as
\begin{equation} \label{OUprocess}
\tau \dot{z}_{\mu}=-z_{\mu}+ \beta \varphi_{\mu}+\sqrt{2 \tau}\zeta_{\mu}(t),
\end{equation}
where $-z_{\mu}$ represents a leakage term, $\varphi_{\mu}$ denotes the input due to the state of the visible layer, $\zeta_{\mu}$ is a white Gaussian noise with zero mean and covariance $\langle \zeta_{\mu}(t) \zeta_{\nu}(t')\rangle = \delta_{\mu ,\nu} \delta(t-t')$, $\tau$ is the typical timescale and $\beta$ tunes the strength of the input fluctuations. In vector notation the field on the analog layer is $\pmb{\varphi}=\tilde{\pmb{X}} \cdot \pmb{\xi} \cdot \pmb{\sigma}/ \sqrt{N}$, or, more explicitly,
\begin{equation}\label{eq:campo1}
\varphi_{\mu}=\frac{1}{\sqrt{N}}\sum_{i=1}^{N}\sum_{\nu=1}^{P} \tilde{X}_{\mu ,\nu} \xi^{\nu}_{i} \sigma_{i} = \frac{1}{\sqrt{N}} \sum_{i=1}^N (c \xi_i^{\mu} + b \xi_i^{\mu+1}) \sigma_i.
\end{equation}
The activity in the digital layer follows a Glauber dynamics as
\begin{equation} \label{Gprocess}
\tau' \langle \dot{m}^{\mu} \rangle_{\xi} = - \langle m^{\mu} \rangle_{\xi}+ \left \langle \xi^{\mu} \tanh \left [ \beta \frac{1}{\sqrt{N}} \phi \right ] \right \rangle_{\xi},
\end{equation}
where the interaction with the hidden layer is encoded by $\pmb{\phi}=\tilde{\pmb{X}} \cdot \pmb{\xi} \cdot \pmb{z}$, that is,
\begin{equation}\label{eq:campo3}
\phi_i = \sum_{\mu=1}^P \sum_{\nu=1}^P \tilde{X}_{\mu,\nu} \xi_{i}^{\nu} z_{\mu} = \sum_{\mu=1}^N (c \xi_i^{\mu} + b \xi_i^{\mu+1}) z_{\mu}.
\end{equation}
The timescale of the analog dynamics (\ref{OUprocess}) is assumed to be much faster than that of the digital one (\ref{Gprocess}), that is $\tau' \gg \tau$.


Since all the interactions are symmetric, it is possible to describe this system through a Hamiltonian formulation:
from the OU process of Eq.~(\ref{OUprocess}) we can write
$$
\tau \dot{z}_{\mu}= - \partial_{z_{\mu}}\tilde{H}(\pmb{z}, \pmb{\sigma}, \pmb{\xi}, \tilde{\pmb{X}}),
$$
being
\be\label{acca}
\tilde{H}(\pmb{z}, \pmb{\sigma}, \pmb{\xi}, \tilde{\pmb{X}}) =  z^2/2 - \beta \sum_{\mu =1}^{P}\varphi_{\mu} z_{\mu}.
\ee
%
%
The partition function $Z_{N}(\beta ;\pmb{\xi}, \tilde{\pmb{X}})$ for such a system then reads off as
\begin{equation}
Z_{N}(\beta ;\pmb{\xi}, \tilde{\pmb{X}})=\sum_{\pmb{\sigma}} \int \prod_{\mu =1}^{P}d\mu(z_{\mu}) e^{- \tilde{H}(\pmb{z}, \pmb{\sigma}, \pmb{\xi}, \tilde{\pmb{X}})},
\end{equation}
where $d\mu(z_{\mu})$ is the Gaussian weight obtained integrating the leakage term in the OU equation.


Now, by performing the Gaussian integration, we get
\begin{equation}
Z_{N, P}(\beta ;\pmb{\xi}, \tilde{\pmb{X}}) = (2 \pi)^{P/2} \sum_{\pmb{\sigma}} e^{\frac{\beta^2}{2} \sum_{\mu =1}^{P}{\varphi_{\mu}^{2}}}=  (2 \pi)^{P/2} \sum_{\pmb{\sigma}} e^{-\frac{\beta^2}{2} H(\pmb{\sigma}, \pmb{\xi}, \tilde{\pmb{X}})},
\end{equation}
where
\begin{equation} \label{hamiltoniana}
H(\pmb{\sigma}, \pmb{\xi}, \tilde{\pmb{X}})=- \pmb{\varphi}^{2}=-\dfrac{1}{N} \pmb{\sigma}^{T} \cdot \pmb{\xi}^{T} \tilde{\pmb{X}}^{T} \tilde{\pmb{X}} \pmb{\xi} \cdot \pmb{\sigma},
\end{equation}
which corresponds to an Hopfield model with patterns $\pmb{\tilde{\xi}} = \tilde{\pmb{X}} \cdot \pmb{\xi}$, under the shift $\beta^2 \rightarrow \beta$.
We then call $\pmb{X}=\tilde{\pmb{X}}^{T} \tilde{\pmb{X}}$ the correlation matrix which is obviously symmetric,\ so that the interactions between the $\sigma$s are symmetric,\ leading to an equilibrium scenario. Using Eq.~\ref{xtilde}, the matrix $\pmb{X}$ is
\begin{equation} \label{x}
X_{\mu ,\nu}=(c^2+b^{2}) \delta_{\mu ,\nu} + cb (\delta_{\mu ,\nu +1} + \delta_{\mu -1 ,\nu }),
\end{equation}
and we can fix $b^2 + c^2 =1$ and $bc=a$, to recover the coupling in Eq.~\ref{eq:J_leticia}.
It is easy to see that, as long as $b, c \in \mathbb{R}$, $a \leq 1/2$.
In general, with some algebra, we get
\begin{eqnarray}
c = \pm \frac{1}{2} (\sqrt{1+2a}  \pm \sqrt{1-2a} ),\\
b = \pm \frac{1}{2} (\sqrt{1+2a}  \mp \sqrt{1-2a} ),
\end{eqnarray}
therefore, the product $\pmb{\tilde{X}} \cdot \pmb{\xi}$ appearing in both fields $\varphi_{\mu}$ (see Eq.~\ref{eq:campo1}) and $\phi_{\mu}$ (see Eq.~\ref{eq:campo3}), turns out to be
\begin{equation}
(\pmb{\tilde{X}} \cdot \pmb{\xi})_{\mu,i} =  \pm \frac{1}{2} [\sqrt{1+2a} (\xi_i^{\mu} + \xi_i^{\mu+1}) \pm \sqrt{1-2a} (\xi_i^{\mu} - \xi_i^{\mu+1})].
\end{equation}
Thus, when $a \leq 1/2$, $(\pmb{\tilde{X}} \cdot \pmb{\xi})_{\mu,i} \in \mathbb{R}, \forall \mu, i$, while for $a > 1/2$, $(\pmb{\tilde{X}} \cdot \pmb{\xi})_{\mu,i}$ can be either real or pure imaginary, according to whether the $\mu$-th entry and the following $\mu+1$-th are aligned or not.

Having described the behavior of the fields, we can now deepen our investigation on the dynamics of the Boltzman machine  underlying our generalized Hopfield model.
\newline
Let us write down explicitly the two coupled stochastic Langevin equations (namely one OU process for the hidden layer, and one Glauber process for the Hopfield neurons) as
\begin{eqnarray} \label{eq:sis1}
    \tau  \dot{z_\mu} &=&-z_\mu +\frac{\beta}{\sqrt{N}}\sum_i (c\xi_i^\mu + b\xi_i^{\mu+1})\sigma_i \\
    \label{eq:sis2}
    \tau' \langle \dot{m_\mu} \rangle &=&- \langle m_\mu \rangle_{\xi} +\left\langle \xi^\mu \tanh[\beta \frac{1}{\sqrt{N}}\sum_\nu^P z_\nu(c\xi^\nu + b\xi^{\nu+1})] \right\rangle_{\xi}.
\end{eqnarray}
Note that by assuming thermalization of the fastest variables with respect to the dynamical evolution of the magnetizations, namely requiring $\dot{z}_{\mu}=0$, we can use Eq.~\ref{eq:sis1} to explicit the term $z_{\nu}$ in the argument of the hyperbolic tangent in Eq.~\ref{eq:sis2}, hence recovering the self-consistencies of Eq.~(\ref{emmes eq}), (see also Appendix C).

Assuming that the two time scales belong to two distinct time sectors, it is possible to proceed in the opposite way, that is
\be
\langle \dot{m_\mu} \rangle \cong0\Rightarrow \langle m_\mu \rangle=\left\langle \xi^\mu \tanh[\beta \frac{1}{\sqrt{N}}
\sum_\nu^P z_\nu(c\xi^\nu + b\xi^{\nu+1})] \right\rangle_{\xi}.
\ee
For the sake of simplicity let us deal with the $P=2$ case, being the generalization to the case $P>2$ straightforward.
Linearization implies
\begin{eqnarray} \label{eq:primosistema_1}
   \tau \dot{z_1} &=& -z_1+\left\langle(c\xi^1+b\xi^2)\beta^2[z_1(c\xi^1+b\xi^2)+z_2(c\xi^2+b\xi^1] \right\rangle_{\xi}, \\
    \label{eq:primosistema_2}
   \tau \dot{z_2} &=& -z_2+\left\langle(c\xi^2+b\xi^1)\beta^2[z_1(c\xi^1+b\xi^2)+z_2(c\xi^2+b\xi^1] \right\rangle_{\xi},
\end{eqnarray}
which, recalling that $c^2+b^2=1$ and $cb=a$, turn out to be
\begin{eqnarray}
    \tau \dot{z_1} &=& z_1\left [ -1+(1-d)\beta^2 \right] +z_2 \left [ 2a(1-d)\beta^2 \right] ,\\
    \tau \dot{z_2} &=& z_2 \left [ -1+(1-d)\beta^2 \right]  +z_1 \left [ 2a(1-d)\beta^2 \right].
\end{eqnarray}
It is convenient to rotate the plane variables $z_1,z_2$ and define
\begin{eqnarray}
   x&=&z_1+z_2 \\
   y&=&z_1-z_2,
\end{eqnarray}
such that Eqs.~\ref{eq:primosistema_1} and \ref{eq:primosistema_2} can be restated as
\begin{eqnarray}
   \dot{x} &=& -x\left[\frac{1}{\tau}-\frac{(1-d)\beta^2 (c+b)^2}{ \tau} \right]\\
   \dot{y} &=& -y\left[\frac{1}{\tau}-\frac{(1-d)\beta^2 (c-b)^2}{ \tau} \right],
\end{eqnarray}
which, in terms of the parameter $a$, are
\begin{eqnarray}
   \dot{x}&=& -x \left [\frac{1}{\tau}-\frac{(1-d)\beta^2 (1+2a)}{\tau} \right]\\
   \dot{y}&=& -y \left [\frac{1}{\tau}-\frac{(1-d)\beta^2 (1-2a)}{\tau} \right],
\end{eqnarray}
whose solution is
\be
\left\{
  \begin{array}{ll}
   x(t)= x(0)e^{Y_x t} = x(0)\exp\big[\frac{-t}{\tau}(1-(1-d)\beta^2(1+2a))\big]\\
   y(t)= y(0)e^{Y_y t} = y(0)\exp\big[\frac{-t}{\tau}(1-(1-d)\beta^2(1-2a))\big].
  \end{array}
\right.
\ee
The Lyapunov exponents of the dynamical system $Y_x,Y_y$ turn out to be
\begin{eqnarray}
Y_x &=& -\frac{1}{\tau}\left[ 1 - \beta^2(1-d)(1+2a) \right ],\\
Y_y &=& -\frac{1}{\tau}\left[ 1 - \beta^2(1-d)(1-2a) \right ].
\end{eqnarray}
This dynamic scenario can be summarized as follows: If the noise level is high $(\beta \ll 1)$, the dynamics is basically quenched on its fixed points $x=0,y=0$ and the corresponding Hopfield model is in the ergodic phase. If the noise level is reduced below the critical threshold, then two behavior may appear: If $a \leq 1/2$ both $x$ and $y$ increase, which means that only one $z$ variable is moving away from its trivial equilibrium state (this corresponds to a retrieval of a single pattern in the generalized Hopfield counterpart); if $a > 1/2$, $x$ increases while $y$ points to zero, which means that both the variables $z_1,z_2$ are moving away from their trivial equilibrium values (this corresponds to a correlated retrieval in the generalized Hopfield counterpart).

Switching to the original variables we get
\begin{eqnarray}\nonumber
  z_1(t) &=& \exp[ \frac{-t}{\tau}(1-(1-d)\beta)](z_1(0)\cosh[\frac{t(1-d)\beta 2a}{\tau}]+z_2(0)\sinh[\frac{t(1-d)\beta 2a}{\tau}]),\\ \nonumber
  z_2(t) &=& \exp[ \frac{-t}{\tau}(1-(1-d)\beta)](z_1(0)\sinh[\frac{t(1-d)\beta 2a}{\tau}]+z_2(0)\cosh[\frac{t(1-d)\beta 2a}{\tau}]).
\end{eqnarray}
Again, Lyapunov exponents describe a dynamics in agreement with the statistical mechanics findings.
%
%

\section{Discussion} \label{sec:discussion}
While technology becomes more and more automatized, our need for a systemic description of cybernetics, able to go over the pure mechanicistic approach, gets more urgent. Among the several ways tried in this sense, neural networks, with their feedback loops among neurons, the multitude of their stable states and their stability under attacks (being the latter noise, dilution or various perturbations), seem definitely promising and worth being further investigated.

Along this line, in this work we considered a complex perturbation of the paradigmatic Hopfield model, by assuming correlation among patterns of stored information \emph{and} dilution in pattern entries. First, we reviewed and deepened both the limiting cases, corresponding to a Hopfield model with correlated attractors (introduced and developed by Amit, Cugliandolo, Griniatsly and Tsodsky \cite{Griniasty-1993NeurComp,Cugliandolo-1993NeurComp,Cugliandolo-1994JPA}) and to a Hopfield model with diluted patterns (introduced by some of us \cite{Agliari-2012lett,Agliari-2012sub}). The general case, displaying a correlation parameter $a > 0$ and a degree of dilution $d>0$, has been analyzed from different perspectives obtaining a consistent and broad description. In particular, we showed that the system exhibits a very rich behavior depending qualitatively on $a$, on $d$ and on the noise $T$: in the phase space there are regions where the pure-state ansatz is recovered, others where several patterns can be retrieved simultaneously and such parallel retrieval can he highly hierarchical or rather homogeneous or even symmetric.

Further, recalling that interactions among spins are symmetric and therefore a Hamiltonian description is always achievable, we can look at the system as the result of marginalization of a suitable (restricted) Boltzman Machine made of by two layers (a visible, digital layer built of by the Hopfield neurons and a hidden, analog layer made of by continuous variables) interconnected by a passive layer of bridges allowing for pattern correlations.
In this way the dynamics of the system can as well be addressed.


\section*{Appendices}

{\bf Appendix A.} - In this Appendix we provide some insights into the shape of the attractors emerging for the correlated model in the noiseless case.
We recall for consistency the coupling
\begin{equation}\label{eq:J_leticia2}
J_{ij} = \frac{1}{N} \sum_{\mu=1}^P [\xi_i^{\mu} \xi_j^{\mu} + a(\xi_i^{\mu+1} \xi_j^{\mu} +\xi_i^{\mu-1} \xi_j^{\mu} ) ],
\end{equation}
where the pattern matrix $\pmb{\xi}$ is quenched.
Due to the definition above, magnetizations are expected to reach a hierarchical structure, where the largest one, say $m^1$, corresponds to the stimulus
and the remaining are symmetrically decreasing, say
\begin{equation}\label{eq:series}
m^1 \geq m^2=m^P \geq ... \geq m^{(P+1)/2} =m^{(P+1)/2+1},
\end{equation}
where we assumed $P$ as odd. The distance between the pattern $\mu$ and the stimulated pattern is $k(\mu,P)=\min[\mu-1,P-(\mu-1)]$.


Moreover, each pattern $\mu$ determines a field $h^{\mu}$, which tends to align the $i$-th spin with $\xi_i^{\mu}$. The field reads off as
\begin{equation}\label{eq:field_mu}
h^{\mu} = m^{\mu} + a (m^{\mu+1} + m^{\mu-1}).
\end{equation}
At zero fast noise we have that
\begin{equation}
\sigma_i =  \textrm{sign}(\varphi_i) = \textrm{sign} \left [ \sum_{\mu=1}^P (\xi_i^{\mu} h^{\mu}) \right].
\end{equation}
Due to Eqs.~(\ref{eq:series}) and (\ref{eq:field_mu}), the first pattern is likely to be associated to a large field and therefore to determine the sign of the overall sum appearing in Eq.~(\ref{eq:field_mu}). On the other hand, patterns with $\mu$ close to $(P+1)/2$ are unlikely to give an effective contribution to $\varphi_i$ and therefore to align the corresponding spins. Indeed, the field $h_{\nu} \xi_i^{\nu}$ may determine the sign of $\varphi_i$ for special arrangements of the patterns $\mu$ corresponding to smaller distance, i.e. $k(\mu,P) < k(\nu,P)$. More precisely, their configuration must be staggered, i.e., under gauge symmetry, $\xi_i^1 = + 1$, $\xi_i^2=\xi_i^P=-1$, $\xi_i^3 = \xi_i^{P-1}=+1, ..., \xi_i^{\nu-1}= \xi^{P-\nu+3}$. By counting such configurations one gets $m^{\nu}$.


With some abuse of language, in the following we will denote with $m_k$ the Mattis magnetization corresponding to patterns at a distance $k$ from the first one. For simplicity, we also assume $P$ small such that $m_k \neq 0, \forall k$.

%
Then, it is easy to see that, over the $2^P$ possible pattern configurations, those which effectively contribute to $m_{(P-1)/2}$ are only $4$. In fact,
it must be $\xi^{(P+1)/2}=\xi^{(P+1)/2+1} = +1 (-1)$ and all the remaining must be staggered; therefore, $m_{(P-1)/2}=4/2^P = 2^{2-P}$.

As for $m_{(P-3)/2}$, contributes come from configurations where the patterns corresponding to $\mu<(P-1)/2$ are staggered. Such configurations are $2^{4}$, but we need to exclude those which are actually ruled by the farthest patterns, which are $4$, hence, the overall contribute is $16-4=12$ and $m_{(P-3)/2}=12/2^P = 3 \times 2^{2-P}$.


We can proceed analogously for the following contributes. In general, by denoting with $c_k$ the $k$-th contribute, one has the following recursive expression
\begin{equation}\label{eq:recursivity}
c_{k-1}=2^{2k} - c_{k}
\end{equation}
with $c_{(P-1)/2}=4$ and $k<(P-1)/2$. For the last contribute, one has $c_{k-1}=2^{2k-1} - c_{k}$, because the last pattern has no ``twin''.

Applying this result we get
\begin{eqnarray}
\pmb{m}&=&\frac{1}{2}(1,1,1), \; \textrm{for} \; P=3\\
\pmb{m}&=&\frac{1}{8}(5,3,1,1,3), \; \textrm{for} \; P=5\\
\pmb{m}&=&\frac{1}{32}(19,13,3,1,1,3,13), \; \textrm{for} \: P=7\\
\pmb{m}&=&\frac{1}{128}(77,51,13,3,1,1,3,13,51), \; \textrm{for} \; P=9,
\end{eqnarray}
consistently with \cite{Cugliandolo-1993NeurComp,Cugliandolo-1994JPA}.

Let us now consider the case $P=11$; following the previous machinery we get $m=\frac{1}{2^9} (307,205,51,13,3,1,1,3,13,51,205)$. However, such state is not stable over the whole range of $a$. In fact, by requiring that the field due to the farthest pattern is larger than the field generated by the staggered configuration of patterns we get
\begin{equation}
2(2a-1)(-m_6/2 + m_5 - m_4  + m_3 - m_2) \leq m_1
\end{equation}
which implies $a< 23/42 \approx 0.54$.  Hence, from that value of $a$, the previous state is replaced by $m=\frac{1}{128}(77,51,13,3,1,0,0,1,3,13,51)$, which is always stable.
Similarly, for $P=13$, we get a state for $m$ with $m_i > 0, \forall i$, which is stable only when $a<85/164 \approx 0.518$, for larger values of $a$ this is replaced by the state found for $P=11$ and then for the state found for $P=9$.

All these results have been quantitatively confirmed numerically.
We finally notice that for the arguments presented here there is no need for the low storage hypothesis.

\smallskip

{\bf Appendix B.} -In this appendix we want to show that the model is well behaved, namely, that its intensive free energy has a thermodynamic limit that exists and is unique: Despite it may look as a redundant check, we stress that the thermodynamic limit of the high storage Hopfield model (e.g. the $\alpha > 0$ case) is still lacking, hence rigorous results on its possible variants still deserve some interest.
\newline To obtain the desired result, our approach follows two steps: first we show, via annealing, that the intensive free energy is bounded in the system size, then we show that it is also super-additive. As a consequence of these two results the statement straightly follows \cite{Ruelle-1999book}.
\newline
Remembering that $F_N(\beta,a,d)=N^{-1}\mathbb{E}\ln Z_N(\beta,a,d)$, where
$$
Z_N(\beta,a,d)=\sum_{\sigma}\exp(-\beta H_N(\sigma;\xi))
$$
is the partition function. Annealing the free energy consists in considering the following bound
\begin{eqnarray}
F_N(\beta,a,d) &=& \left\langle \frac1N \log \sum_{\sigma}e^{-\beta H_N(\sigma;\xi)} \right\rangle \\
&\leq&  \frac1N \log \langle \sum_{\sigma}e^{-\beta H_N(\sigma;\xi)} \rangle \\
&\leq& \frac1N \log  \sum_{\sigma}e^{-\beta  \langle H_N(\sigma;\xi)\rangle},
\end{eqnarray}
where in the last line we used Jensen inequality.
\newline
As a result we get
\begin{eqnarray}
Z_N(\beta,a,d) &=& \sum_{\sigma}e^{ \frac{\beta N}{2} \sum_{\mu}^P \left \{\langle m_{\mu}^2(\sigma) \rangle + a[\langle m_{\mu}(\sigma)m_{\mu+1}(\sigma) \rangle + \langle m_{\mu}(\sigma)m_{\mu-1}(\sigma)\rangle] \right \} } \\
&\leq& 2^N e^{\frac{P}{2}(N-1)\beta(1+2a)(1-d),}
\end{eqnarray}
by which the annealed free energy bound reads off as
\be
F_N(\beta,a,d) \leq \ln 2 + \frac{P}{2}\beta(1+2a)(1-d)\left(1 + \frac1N \right),
\ee
such that the annealed free energy is $F_A(\beta,a,d)= \ln 2 + P\beta(1+2a)(1-d)/2$.
\newline
Let us move over toward proving the super-additivity property and  consider two systems independent of each other and with respect to to the original $N$-neurons model, and made of respectively by $N_1$ and $N_2$ neurons, such that $N=N_1+N_2$.
\newline
In complete analogy with the original system we can introduce
$$
m_{\mu}^{(1)}= \frac{1}{N_1}\sum_i^{N_1} \xi_i^{\mu}\sigma_i^{(1)},\ \ m_{\mu}^{(2)}= \frac{1}{N_2}\sum_i^{N_2} \xi_i^{\mu}\sigma_i^{(2)},
$$
and note that the original Mattis magnetizations are linear combinations of the sub-systems counterparts such that
$$
m_{\mu} = \frac{N_1}{N}m_{\mu}^{(1)}+\frac{N_2}{N}m_{\mu}^{(2)}.
$$
Since the function $x \to x^2$ is convex (and the translation $x \to \tilde{x}x$ innocent) we have
\begin{eqnarray}
Z_N(\beta,a,d) \nonumber
&\leq& \sum_{\sigma}e^{\beta  N_1 \sum_{\mu}^P \left \{  m_{\mu}^{(1),2}(\sigma) + a \left [m_{\mu}^{(1)}(\sigma) m_{\mu+1}^{(1)}(\sigma) + m_{\mu}^{(1)}(\sigma) m_{\mu-1}^{(1)}(\sigma) \right] \right \} } \\ \nonumber
&\cdot& e^{\beta  N_2 \sum_{\mu}^P \left\{  m_{\mu}^{(2),2}(\sigma) + a \left [m_{\mu}^{(2)}(\sigma) m_{\mu+1}^{(2)}(\sigma)+ m_{\mu}^{(2)}(\sigma) m_{\mu-1}^{(2)}(\sigma) \right] \right \} } \\
&=& Z_{N_1}(\beta,a,d)Z_{N_2}(\beta,a,d),
\end{eqnarray}
by which the free energy density $F_N(\beta,a,d)$ is shown to be sub-additive as
$$
N F_N(\beta,a,d) \geq N_1 F_{N_1}(\beta,a,d) + N_2 F_{N_2}(\beta,a,d).
$$
As the free energy density is sub-additive and it is limited (and this is an obvious consequence of the annealed bound), the infinite volume limit exists and is unique and equal to its sup over the system size $\lim_{N \to \infty} F_N(\beta,a,d) = \sup_N F_N(\beta,a,d) = F(\beta,a,d)$.

\smallskip

{\bf Appendix C.}
In this Appendix we outline the statistical mechanics calculations that brought to the self consistency used in the text (eq. \ref{eq:selfcons_a}). Our calculations are based on the Hamilton-Jacobi interpolation technique \cite{Guerra-2001FIC}\cite{Barra-2008JSP}\cite{Genovese-2009JMP}. This appendix aims two different targets. From one side it outlines the physics of the model and describes it through the self-consistent equation; from the other side it develops a novel mathematical technique able to solve this kind of statistical mechanics problems.
\newline
In a nutshell the idea is to think at $\beta$ as a ''time-variable" and to introduce $P$ ficticious axes $x_{\mu}$, meant as ''space-variables", then, within an Hamilton-Jacobi framework, the free energy with respect to these Euclidean coordinates, is shown to play the role of the Principal Hamilton Function, whose solution can then be extrapolated from classical mechanics.
\newline
Our generalization of the Hopfield model is described by the Hamiltonian:
\begin{equation}
H_N(\bold{\sigma},\bold{\xi})=- \frac{1}{2N}\sum_{i, j}^N\sigma_i \sigma_j \sum_{\mu,\nu}^P \xi_i^\mu X_{\mu,\nu} \xi_j^\nu,
\end{equation}
as discussed in the text (see Sects.~$2$ and $3$).
\newline
The $N$-neuron partition function $Z_N(\beta,a,d)$ and the free energy $F(\beta,a,d)$ can be written as
\begin{eqnarray}
Z_N(\beta,a,d) &=& \sum_{\sigma}\exp\left[-\beta H_N(\bold{\sigma},\bold{\xi}) \right], \\
F(\beta,a,d) &=&  \lim_{N\to \infty} \frac1N \langle \log Z_N(\beta,a,d)\rangle,
\end{eqnarray}
where $\langle . \rangle$ again denotes the full averages over both the distribution of the quenched patterns $\xi$ and the Boltzman weight (for the sake of clearness, let us stress that the factor $B(\beta,a,d) = \exp[-\beta H_N(\bold{\sigma},\bold{\xi}) ]$ is termed Boltzman factor).

As anticipated, the idea of the Hamilton-Jacobi interpolation is to enlarge the ''space of the parameters" by introducing a $P+1$ Euclidean structure (where $P$ dimensions are of space type and mirrors the $P$ Mattis magnetization, while the remaining one is of time type and mirrors the temperature dependence) and to find a general solution for the free energy in this space thanks to techniques stemmed from classical mechanics. The statistical mechanics free energy will then be simply this extended free energy evaluated in a particular point of this larger space.
Analogously, the average $\langle.\rangle_{(\bold{x},t)}$ extends the ones earlier introduced by accounting for this generalized Boltzmann factor and will be denoted by $\langle.\rangle$, wherever evaluated in the sense of statistical mechanics.
\newline
The ``Euclidean" free energy for $N$ neurons, namely $F_N(t,\bold{x})$, can then be written in vectorial terms as
\begin{equation}
F_N(t,\bold{x})=\frac{1}{N} \left\langle \ln\left \{ \sum_{\sigma}\exp \left[ \frac{-t}{2N}( \bold{\xi} \bold{\sigma}, X \bold{\xi}\bold{\sigma} ) +
(\bold{x}, \bold{\xi} \bold{\sigma}) \right] \right \}\right\rangle.
\end{equation}
The matrix $\pmb{X}$ can be diagonalized trough $X = U^{\dagger}D U$, where $U$ and $U^{\dagger}$ are (unitary) rotation matrices and $D$ is the diagonal expression, such that
\begin{eqnarray}
F_N(t,\bold{x})&=&\frac1N \left\langle \ln \sum_{\sigma}\exp \left[ -\frac{t}{2N}(\bold{\xi} \bold{\sigma}, X \bold{\xi} \bold{\sigma})+ (\bold{x}, \bold{\xi}\bold{\sigma}) \right]\right\rangle,\\
&=& \frac1N \left\langle \ln \sum_{\sigma} \exp \left[ -\frac{t}{2N}(\sqrt{D}U \bold{\xi}\bold{\sigma}, \sqrt{D}U \bold{\xi}\bold{\sigma}) + (\sqrt{D}^{-1}U^{\dagger}\bold{x}, \sqrt{D}U \bold{\xi}\bold{\sigma})\right]\right\rangle, \nonumber
\end{eqnarray}
as $(\bold{\xi}\bold{\sigma}, X \bold{\xi} \bold{\sigma})= (\bold{\xi}\bold{\sigma}, U^{\dagger}D U \bold{\xi} \bold{\sigma})=(\sqrt{D}U \bold{\xi} \bold{\sigma},\sqrt{D}U \bold{\xi}\bold{\sigma})$ and $(\bold{x},\bold{\xi} \bold{\sigma})=(\bold{x}, U^{\dagger}U \bold{\xi}\bold{\sigma})=(\sqrt{D}^{-1}U \bold{x}, \sqrt{D} U \bold{\xi} \bold{\sigma})$.
If we switch to the new variables $\bold{\tilde{\xi}}=\sqrt{D}U\bold{\xi}$ and $\bold{\tilde{x}}=\sqrt{D}^{-1}U \bold{x}$ we can write the Euclidean free energy in a canonical form as
\begin{eqnarray}
F_N(t,\bold{x})&=&\frac1N \left\langle \ln \sum_{\sigma} \exp\Big( -\frac{t}{2N}(\tilde{\bold{\xi}}\bold{\sigma},\bold{\tilde{\xi}}\bold{\sigma})+(\bold{\tilde{x}},\bold{\tilde{\xi}}
\bold{\sigma})\Big)\right\rangle \\
&=& \frac1N \left\langle \ln \sum_{\sigma} \exp\Big( \sum_{ij}^N \sigma_i\sigma_j \sum_{\mu}\bold{\tilde{\xi}}_i^{\mu}\bold{\tilde{\xi}}_j^{\mu} + \sum_{\mu} \bold{\tilde{x}}_{\mu}\sum_j^N \bold{\tilde{\xi}}_j^{\mu}\sigma_j \Big)\right\rangle.
\end{eqnarray}
Thus, we write the $(\bold{x},t)$-dependent Boltzmann factor as
\begin{equation}
B_N(\bold{x},t)=\exp\left[ \frac{-t}{2N}(\tilde{\bold{\xi}}\bold{\sigma},\bold{\tilde{\xi}}\bold{\sigma})+
(\bold{\tilde{x}},\bold{\tilde{\xi}}
\bold{\sigma}) \right ],
\end{equation}
remembering that $B_N(\bold{x},t)$ matches the classical statistical mechanics factor for $t=-\beta$ and $x_\mu=0$ $\forall \mu$, as even a visual check can immediately confirm.

Now, let us consider the derivative of the free energy with respect to each dimension (i.e., $t, \tilde{x}_{\mu}$):
\begin{eqnarray}
\partial_t F_N(t,\bold{x}) &=&-\frac{1}{2}\sum_{\mu=1}^P \langle (\tilde{m}_\mu)^2 \rangle_{(\bold{\tilde{x}},t)},\\
\partial_{\tilde{x}_\mu}F_N(t,\bold{\tilde{x}}) &=& \langle \tilde{m}_\mu  \rangle_{(\bold{\tilde{x}},t)}.
\end{eqnarray}

We notice that the free energy implicitly acts as a Principal Hamilton Action if we introduce the potential
\begin{equation}
V_N(t,\bold{\tilde{x}})=\frac{1}{2}\sum_{\mu=1}^P \left[\langle (\tilde{m}_\mu)^2  \rangle - \langle \tilde{m}_\mu  \rangle^2\right].
\end{equation}
In fact, we can write the Hamilton-Jacobi equation for the $F_N$ action as
\begin{equation}\label{HJ}
\partial_t F_N(t,\bold{\tilde{x}})+\frac{1}{2}\sum_\mu \bigg(\partial_{\tilde{x}_\mu}F_N(t,\bold{\tilde{x}}) \bigg)^2+V_N(t,\bold{\tilde{x}})=0.
\end{equation}
Interestingly, the potential is the sum of the variances of the order parameters and we know from Central Limit Theorem argument that in the thermodynamic limit they must vanish, one by one, and, consequently, $\lim_{N \to \infty}V_N(t,\bold{\tilde{x}})=0$.
Such self-averaging property play a key role in our approach as, in the thermodynamic limit, the motion turns out to be free. Moreover, as shown in Appendix A, the limit
\be
F(t,\bold{\tilde{x}})=\lim_{N\to\infty}F_N(t,\bold{\tilde{x}}),
\ee
exists, and $F(t,\bold{\tilde{x}})$ can then be obtained solving the free-field Hamilton-Jacobi problem as
\begin{equation}\label{HJfree}
\partial_t F(t,\bold{\tilde{x}})+\frac{1}{2}\sum_\mu \bigg(\partial_{\tilde{x}_\mu}F(t,\bold{\tilde{x}}) \bigg)^2=0.
\end{equation}
From standard arguments of classical mechanics, it is simple to show that the solution for the Principal Hamilton Function, i.e. the free energy, is
the integral of the Lagrangian over time plus the initial condition (which has the great advantage of being a trivial one-body calculation as $t=0$ decouples the neurons).
More explicitly,
\begin{equation}
F(t,\bold{\tilde{x}})= F(t_0,\bold{\tilde{x}}_0)+\int_0^t dt' \mathcal{L}(t',\bold{\tilde{x}}),
\end{equation}
where the Lagrangian can be written as
\begin{equation}
\mathcal{L}(t,\bold{\tilde{x}})= \frac12 \sum_{\mu=1}^P \Big( \partial_{x_{\mu}} F(t,\bold{\tilde{x}}) \Big)^2 = \frac{1}{2}\sum_\mu \langle \tilde{m}_\mu  \rangle^2.
\end{equation}
Having neglected the potential, the motion must be constrained in straight hyperplanes, and the Cauchy problem is
\begin{equation}
\left\{
 \begin{aligned}
 & t_0=0 \\
 & \tilde{x}_\mu= \tilde{x}_\mu^0 + t \langle \tilde{m}_\mu  \rangle
\end{aligned}
 \right.
 \end{equation}
We can now write the solution more explicitly as
\begin{eqnarray}
F(t,\bold{\tilde{x}}) &=& F(0,\bold{\tilde{x}}_0)+\int dt' \mathcal{L}(t',\bold{\tilde{x}}) \\ \nonumber
&=&\frac{t}{2}\sum_\mu \langle \tilde{m}_\mu  \rangle^2 + \lim_{N \to \infty }\frac{1}{N}\sum_j^N \ln \left [\sum_\sigma \exp \left ( \sigma_j \sum_\mu \tilde{x}_\mu^0 \tilde{\xi}_j^\mu \right) \right ]\\ \nonumber
&=&\ln 2+\frac{t}{2}\sum_\mu \langle \tilde{m}_\mu  \rangle^2+ \left \langle \ln \left \{ \cosh \left [\sum_\mu (\tilde{x}_\mu -t \langle \tilde{m}_\mu\rangle)\tilde{\xi}^\mu \right] \right \} \right \rangle.
\end{eqnarray}

As a consequence, the free energy of this generalization of the Hopfield model can be written by choosing $t=-\beta$ and $\tilde{x}_{\mu}=0$ for all the spatial dimensions, so to have

\begin{equation}
F(\beta,a,d)=\ln 2-\frac{\beta}{2}\sum_\mu \langle \tilde{m}_\mu  \rangle^2
+\langle \ln \big[\cosh[\beta \sum_\mu \langle \tilde{m}_\mu\rangle \tilde{\xi}^\mu] \big]\rangle.
\end{equation}

We can proceed to extremization, namely $\partial_{\tilde{m}_\mu}F(\beta,a,d)=0$ to get
\begin{equation}
\langle \tilde{m}_\mu\rangle = \langle \tilde{\xi}^\mu \tanh[\beta\sum_\mu \langle \tilde{m}_\mu \rangle
\tilde{\xi}^\mu]     \rangle,
\end{equation}
which, turning to the original variables, can be written as
\begin{eqnarray}
F(\beta,a,d)&=& \ln 2 -\frac{\beta}{2}\sum_{\mu}^P \langle m_{\mu}^2 \rangle + \langle \ln \cosh \beta ( \xi, X m )\rangle,\\
\langle m_{\mu}\rangle &=& \left \langle \xi_{\mu} \tanh\left[ \beta (\xi, Xm) \right ] \right \rangle.
\end{eqnarray}
which are the equations that have been used trough the text.
\smallskip
For the sake of clearness, the expression
for the Mattis magnetizations
\be
m_\mu= \left \langle \xi^\mu \tanh \left [ \frac{\beta}{P}\sum_\nu z_\nu(c\xi^\nu+b\xi^{\nu+1}) \right]  \right \rangle
\ee
is written extensively for $P=2$, namely
\begin{eqnarray}\nonumber
m_1 &=& \frac{(1-d)^2}{2}\tanh[\frac{\beta}{2}(z_1+z_2)(c+b)]+ \\
&+&\frac{(1-d)^2}{2}\tanh[\frac{\beta}{2}(z_1-z_2)(c-b)]+d(1-d) \tanh \left [\frac{\beta}{2}(z_1 c+z_2 b) \right],\\
m_2 &=& \frac{(1-d)^2}{2}\tanh[\frac{\beta}{2}(z_1+z_2)(c+b)]+\frac{(1-d)^2}{2} \tanh[\frac{\beta}{2}(z_1-z_2)(c-b)]+ \nonumber \\
&+& d(1-d) \tanh[\frac{\beta}{2}\frac{1}{2}((z_1+z_2)(c+b)-(z_1-z_2)(b-c)  )],
\end{eqnarray}

and for $P=3$, namely

\begin{eqnarray}
\nonumber
m_{1}&=&
  d^{2}(1-d) \tanh \beta \left(m_{1}+a(m_{2}+m_{3}) \right) \\
  \nonumber
&+&  \dfrac{d(1-d)^{2}}{2} \tanh \beta \left(m_{1}+m_{3}+a(m_{1}+2m_{2}+m_{3}) \right) \\
\nonumber
&+&  \dfrac{d(1-d)^{2}}{2} \tanh \beta \left(m_{1}-m_{3}+a(m_{3}-m_{1}) \right) \\
\nonumber
&+&   \dfrac{d(1-d)^{2}}{2} \tanh \beta \left(m_{1}+m_{2}+a(m_{1}+m_{2}+2m_{3}) \right)  \\
   \nonumber
&+& \dfrac{d(1-d)^{2}}{2} \tanh \beta \left(m_{1}-m_{2}+a(m_{2}-m_{1}) \right)  \\
   \nonumber
&+& \dfrac{(1-d)^{3}}{4} \tanh \beta \left(m_{1}-m_{2}-m_{3}-2 a m_{1} \right)  \\
   \nonumber
&+&   \dfrac{(1-d)^{3}}{4} \tanh \beta \left(m_{1}+m_{2}-m_{3}+2 a m_{3} \right)  \\
   \nonumber
 &+&  \dfrac{(1-d)^{3}}{4} \tanh \beta \left(m_{1}-m_{2}+m_{3}+2 a m_{2} \right)  \\
    \nonumber
 &+&  \dfrac{(1-d)^{3}}{4} \tanh \beta \left(m_{1}+m_{2}+m_{3}+2a(m_{1}+m_{2}+m_{3}) \right) ,
\end{eqnarray}
as for $m_{2}$ and $m_{3}$, they can be obtained through direct permutation $m_1 \to m_2 \to m_3 \to m_1$.

\begin{center}
***
\end{center}

\smallskip

\noindent
This work is supported by FIRB grant RBFR08EKEV. \\
Sapienza Universita' di Roma and INFN are acknowledged too for partial financial support. \\
The authors are grateful to Ton Coolen and Francesco Moauro for useful discussions.


\begin{thebibliography}{9}


\bibitem{Minsky-1969book}
M.~Minsky, S.~Papert, Perceptrons, (enlarged edition, 1988) Edition, MIT Press,
  1969.

\bibitem{Turing-1950Mind}
A.~M. Turing, Computing machinery and intelligence, Mind 49 (1950) 433.

\bibitem{neumann-book}
S.~J. Heims, John von Neumann and Norbert Wiener: From Mathematics to the
  Technologies of Life and Death, MIT Press, 1980.

\bibitem{Kinchin-1957book}
A.~Kinchin, Mathematical foundation of information theory, Dover Publications,
  1957.

\bibitem{hopfield-pnas}
J.~Hopfield, Neural networks and physical systems with emergent collective
  computational abilities, Proc. Natl. Acad. Sc. USA 79 (1982) 2554--2558.

\bibitem{Amit-2003book}
D.~Amit, Modeling Brain Function, Cambridge University Press, 1989.

\bibitem{Jaynes-1957PR}
E.~Jaynes, Information theory and statistical mechanics, Physical Review 106
  (1957) 620.

\bibitem{Kinchin-1949book}
A.~I. Kinchin, Mathematical foundation of statistical mechanics, Dover
  publications, 1949.

\bibitem{Mezard-1987book}
M.~M\'ezard, G.~Parisi, M.~A. Virasoro, Spin glass theory and beyond, World
  Scientific, Singapore, 1987.

\bibitem{Barra-2008JSP}
A.~Barra, The mean field ising model trough interpolating techniques, Journal
  of Statistical Physics 132 (2008) 787.

\bibitem{Kistler-2002book}
W.~K. W.~Gerstner, Spiking neuron models: Single neurons, populations,
  plasticity, Cambridge University Press, 2002.

\bibitem{Griniasty-1993NeurComp}
M.~Griniasty, M.~Tsodyks, D.~Amit, Convertion of temporal correlations between
  stimuli to spatial correlations between attractors, Neural Computation 5
  (1993) 1.

\bibitem{Cugliandolo-1993NeurComp}
L.~Cugliandolo, Correlated attractors from uncorrelated stimuli, Neural
  Computation 6 (1993) 220.

\bibitem{Cugliandolo-1994JPA}
L.~Cugliandolo, M.~Tsodyks, Capacity of networks with correlated attractors,
  Journal of Physics A Mathematical and Theoretical 27 (1994) 741.

\bibitem{Agliari-2012lett}
E.~Agliari, A.~Barra, A.~Galluzzi, F.~Guerra, F.~Moauro, Multitasking
  associative networks, submitted.

\bibitem{Agliari-2012sub}
E.~Agliari, A.~Barra, A.~Galluzzi, F.~Guerra, F.~Moauro, Parallel processing in
  immune networks, submitted.

\bibitem{Guerra-2001FIC}
F.~Guerra, Sum rules for the free energy in the mean field spin glass model,
  Fields Institute Communications 30 (2001) 161--170.

\bibitem{Genovese-2009JMP}
G.~Genovese, A.~Barra, A mechanical approach to mean field models, Journal of
  Mathematical Physics 50 (2009) 053303.

\bibitem{Bengio-2009ML}
Y.~Bengio, Learning deep architectures for artificial intelligence, Machine
  Learning 2 (2009) 127.

\bibitem{Barra-2010JSP}
A.~Barra, F.~Guerra, G.~Genovese, The replica symmetric behavior of the
  analogical neural network, Journal of Statistical Physics 140~(4) (2010) 784.

\bibitem{Barra-2012NN}
A.~Barra, A.~Bernacchia, E.~Santucci, P.~Contucci, On the equivalence of
  hopfield networks and boltzmann machines, Neural Networks.

\bibitem{Coolen-2005book}
A.~Coolen, R.~K\"{u}hn, P.~Sollich, Theory of Neural Information Processing
  Systems, Oxford University Press, 2005.

\bibitem{Chang-1988a}
Y.~Miyashita, Neuronal correlate of visual associative long-term memory in the
  primate temporal cortex, Nature 335 (1988) 817.

\bibitem{Chang-1988b}
Y.~Miyashita, H.~Chang, Neuronal correlate of pictorial short-term memory in
  the primate temporal cortex, Nature 331 (1988) 68.

\bibitem{Glauber-1963JMP}
R.~J. Glauber, Time-dependent statistic of the ising model, Journal of
  Mathematical Physics 4 (1963) 294.

\bibitem{Ruelle-1999book}
D.~Ruelle, Statistical mechanics: rigorous results, World Scientific, 1999.

\end{thebibliography}
\end{document}